\documentclass[iop,revtex4,numberedappendix,twocolappendix]{emulateapj}

\usepackage{graphicx}
\usepackage{mathtools}
\usepackage{gensymb}
\usepackage{enumitem}
\usepackage{hyperref}
\usepackage{placeins}
\usepackage{textgreek}
\usepackage{layouts}

\newcommand{\hf}{\hfill\hspace{0pt}}
\newcommand{\ra}[4]{$#1^{\rm h}#2^{\rm m}#3^{\rm s}\mathllap{.}#4$}
\newcommand{\dec}[4]{$#1\degree#2'#3'\mathclap{.}'#4$}

\newcounter{Affiliation}
\newcommand{\aftext}[1]{\refstepcounter{Affiliation}\altaffilmark{\theAffiliation}#1}

\shorttitle{Type~I\lowercase{bn} Supernovae}
\shortauthors{Hosseinzadeh et al.}

\begin{document}

\title{Type~I\lowercase{bn} Supernovae Show Photometric Homogeneity and\\Spectral Diversity at Maximum Light}

\author{
Griffin~Hosseinzadeh\altaffilmark{\ref{af:LCOGT},\ref{af:UCSB}},
Iair~Arcavi\altaffilmark{\ref{af:LCOGT},\ref{af:KITP}},
Stefano~Valenti\altaffilmark{\ref{af:Davis}},
Curtis~McCully\altaffilmark{\ref{af:LCOGT},\ref{af:UCSB}},
D.~Andrew~Howell\altaffilmark{\ref{af:LCOGT},\ref{af:UCSB}},
Joel~Johansson\altaffilmark{\ref{af:Weizmann}},
Jesper~Sollerman\altaffilmark{\ref{af:OKC}},
Andrea~Pastorello\altaffilmark{\ref{af:OAPD}},
Stefano~Benetti\altaffilmark{\ref{af:OAPD}},
Yi~Cao\altaffilmark{\ref{af:Caltech}},
S.~Bradley~Cenko\altaffilmark{\ref{af:Goddard},\ref{af:UMD}},
Kelsey~I.~Clubb\altaffilmark{\ref{af:Berkeley}},
Alessandra~Corsi\altaffilmark{\ref{af:TTU}},
Gina~Duggan\altaffilmark{\ref{af:Caltech}},
Nancy~Elias-Rosa\altaffilmark{\ref{af:OAPD}},
Alexei~V.~Filippenko\altaffilmark{\ref{af:Berkeley}},
Ori~D.~Fox\altaffilmark{\ref{af:STSci}},
Christoffer~Fremling\altaffilmark{\ref{af:OKC}},
Assaf~Horesh\altaffilmark{\ref{af:Caltech}},
Emir~Karamehmetoglu\altaffilmark{\ref{af:OKC}},
Mansi~Kasliwal\altaffilmark{\ref{af:Caltech}},
G.~H.~Marion\altaffilmark{\ref{af:UT}},
Eran~Ofek\altaffilmark{\ref{af:Weizmann}},
David~Sand\altaffilmark{\ref{af:TTU}},
Francesco~Taddia\altaffilmark{\ref{af:OKC}},
WeiKang~Zheng\altaffilmark{\ref{af:Berkeley}},
Morgan~Fraser\altaffilmark{\ref{af:IoA}},
Avishay~Gal-Yam\altaffilmark{\ref{af:Weizmann}},
Cosimo~Inserra\altaffilmark{\ref{af:QUB}},
Russ~Laher\altaffilmark{\ref{af:IPAC}},
Frank~Masci\altaffilmark{\ref{af:IPAC}},
Umaa~Rebbapragada\altaffilmark{\ref{af:JPL}},
Stephen~Smartt\altaffilmark{\ref{af:QUB}},
Ken~W.~Smith\altaffilmark{\ref{af:QUB}},
Mark~Sullivan\altaffilmark{\ref{af:Southampton}},
Jason~Surace\altaffilmark{\ref{af:IPAC}}, and
Przemek~Wo\'{z}niak\altaffilmark{\ref{af:LANL}}
}
\affil{
\aftext{Las Cumbres Observatory, 6740 Cortona Dr Ste 102, Goleta, CA 93117-5575, USA; \href{mailto:griffin@lco.global}{griffin@lco.global}\label{af:LCOGT}}\\
\aftext{Department of Physics, University of California, Santa Barbara, CA 93106-9530, USA\label{af:UCSB}}\\
\aftext{Kavli Institute for Theoretical Physics, University of California, Santa Barbara, CA 93106-4030, USA\label{af:KITP}}\\
\aftext{Department of Physics, University of California, 1 Shields Ave, Davis, CA 95616-5270, USA\label{af:Davis}}\\
\aftext{Department of Particle Physics and Astrophysics, Weizmann Institute of Science, 76100 Rehovot, Israel\label{af:Weizmann}}\\
\aftext{Oskar Klein Centre, Department of Astronomy, Stockholm University, Albanova University Centre, SE-106 91 Stockholm, Sweden\label{af:OKC}}\\
\aftext{INAF-Osservatorio Astronomico di Padova, Vicolo dell'Osservatorio 5, I-35122 Padova, Italy\label{af:OAPD}}\\
\aftext{Cahill Center for Astronomy and Astrophysics, California Institute of Technology, Mail Code 249-17, Pasadena, CA 91125, USA\label{af:Caltech}}\\
\aftext{Astrophysics Science Division, NASA Goddard Space Flight Center, Mail Code 661, Greenbelt, MD 20771, USA\label{af:Goddard}}\\
\aftext{Joint Space-Science Institute, University of Maryland, College Park, MD 20742, USA\label{af:UMD}}\\
\aftext{Department of Astronomy, University of California, Berkeley, CA 94720-3411, USA\label{af:Berkeley}}\\
\aftext{Department of Physics, Texas Tech University, Box 41051, Lubbock, TX 79409-1051, USA\label{af:TTU}}\\
\aftext{Space Telescope Science Institute, 3700 San Martin Dr, Baltimore, MD 21218, USA\label{af:STSci}}\\
\aftext{University of Texas at Austin, 1 University Station C1400, Austin, TX 78712-0259, USA\label{af:UT}}\\
\aftext{Institute of Astronomy, University of Cambridge, Madingley Road, Cambridge CB3 0HA, UK\label{af:IoA}}\\
\aftext{Astrophysics Research Centre, School of Mathematics and Physics, Queen's University Belfast, Belfast BT7 1NN, UK\label{af:QUB}}\\
\aftext{Spitzer Science Center, California Institute of Technology, Pasadena, CA 91125, USA\label{af:IPAC}}\\
\aftext{Jet Propulsion Laboratory, California Institute of Technology, 4800 Oak Grove Dr, Pasadena, CA 91109, USA\label{af:JPL}}\\
\aftext{Department of Physics and Astronomy, University of Southampton, Southampton SO17 1BJ, UK\label{af:Southampton}}\\
\aftext{Space and Remote Sensing, MS B244, Los Alamos National Laboratory, Los Alamos, NM 87545, USA\label{af:LANL}}
}
\slugcomment{Received 2016 August 3; revised 2016 August 29; accepted 2016 August 31}

\begin{abstract}

Type~Ibn supernovae (SNe) are a small yet intriguing class of explosions whose spectra are characterized by low-velocity helium emission lines with little to no evidence for hydrogen. The prevailing theory has been that these are the core-collapse explosions of very massive stars embedded in helium-rich circumstellar material (CSM). We report optical observations of six new SNe~Ibn: PTF11rfh, PTF12ldy, iPTF14aki, iPTF15ul, SN~2015G, and iPTF15akq. This brings the sample size of such objects in the literature to 22. We also report new data, including a near-infrared spectrum, on the Type~Ibn SN~2015U. In order to characterize the class as a whole, we analyze the photometric and spectroscopic properties of the full Type~Ibn sample. We find that, despite the expectation that CSM interaction would generate a heterogeneous set of light curves, as seen in SNe~IIn, most Type~Ibn light curves are quite similar in shape, declining at rates around 0.1~mag~day$^{-1}$ during the first month after maximum light, with a few significant exceptions. Early spectra of SNe~Ibn come in at least two varieties, one that shows narrow P~Cygni lines and another dominated by broader emission lines, both around maximum light, which may be an indication of differences in the state of the progenitor system at the time of explosion. Alternatively, the spectral diversity could arise from viewing-angle effects or merely from a lack of early spectroscopic coverage. Together, the relative light curve homogeneity and narrow spectral features suggest that the CSM consists of a spatially confined shell of helium surrounded by a less dense extended wind.
\smallskip
\end{abstract}


\section{Introduction}

\setcounter{footnote}{20}

\begin{figure}
\includegraphics[width=\columnwidth]{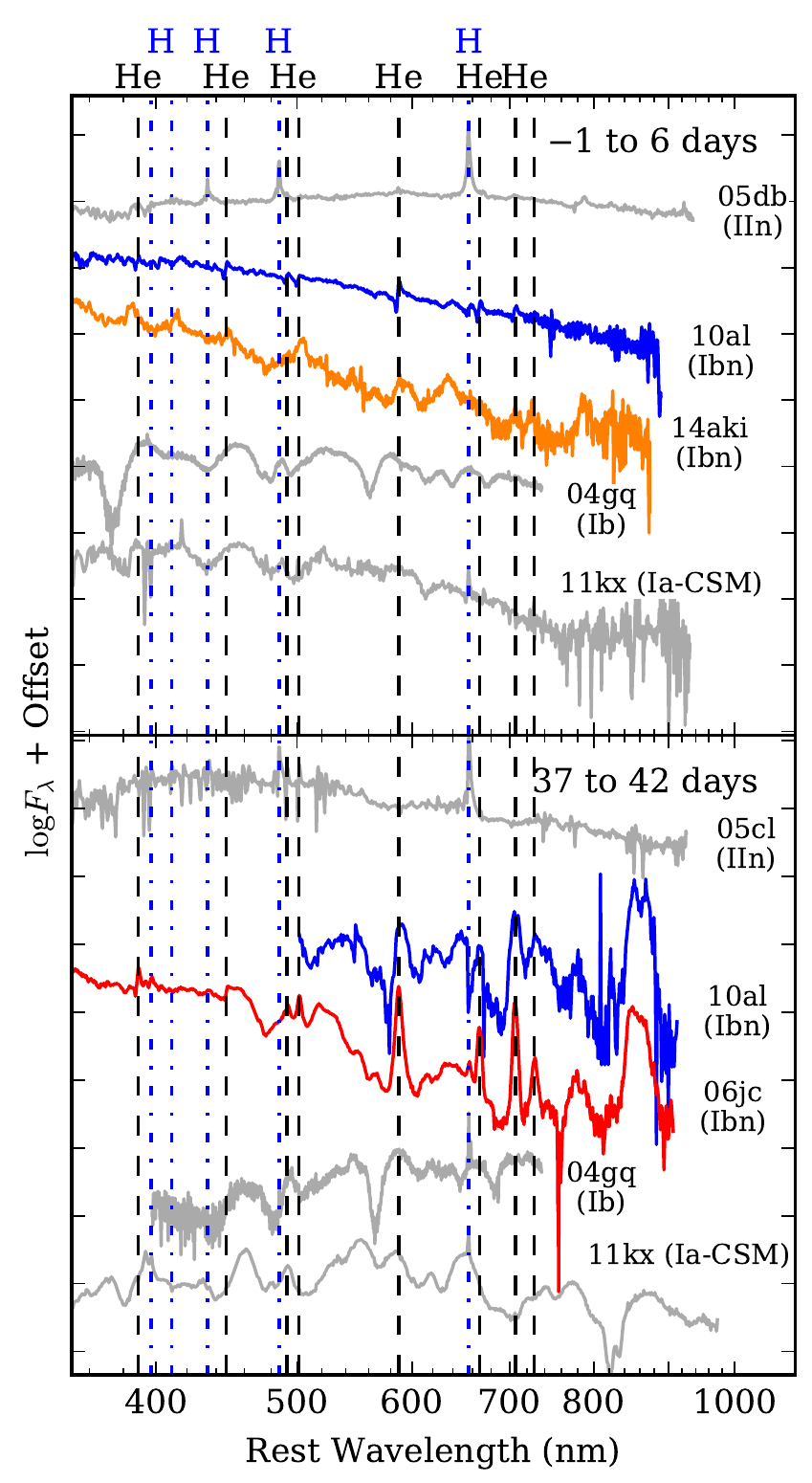}
\caption{Representative spectra of SNe~Ibn compared to those of Type~IIn, Type~Ib, and Type~Ia-CSM supernovae. The upper panel shows spectra near maximum light, while the lower panel shows spectra more than one month later. Spectra of SNe~2005cl and 2005db are from \cite{Kiewe2012}. Spectra of SN~2010al are from \cite{Pastorello4}. The spectrum of SN~2006jc is from \cite{Pastorello2007}. Spectra of SN~2004gq are from \cite{Modjaz2014}. Spectra of PTF11kx are from \cite{Dilday2012}.}
\label{fig:comparespec}
\end{figure}

Supernovae (SNe) strongly interacting with circumstellar material (CSM) provide a unique window into the final evolutionary stages of certain types of massive ($\gtrsim 8 M_\Sun$) stars. For example, observations of CSM interaction probe a star's composition and mass-loss rate immediately prior to its terminal explosion. This can provide insight into the state of the progenitor star in ways that are not available for non-interacting supernovae.

Interaction usually manifests itself as narrow emission lines in spectra of the supernova---broader than lines from H~{\sc ii} regions in the host galaxy but not as broad as lines produced in the ejecta---indicative of the pre-explosion CSM velocity or of the velocity of CSM accelerated by the interaction. Three major spectral classes have been identified (see Figure~\ref{fig:comparespec}). SNe~IIn \citep{Schlegel1990,Filippenko1997}, the most commonly observed interacting supernovae, are core-collapse explosions of hydrogen-rich massive stars (SNe~II) surrounded by hydrogen-rich CSM \citep{Chugai1991}. Type~Ia-CSM supernovae, though similar in appearance to SNe~IIn, are thought to be the explosions of white dwarfs (SNe~Ia) that interact with material stripped from a hydrogen-rich companion \citep[see][and references therein]{Silverman2013}. Finally, the Type~Ibn classification was first proposed by \cite{Pastorello2007} to describe several hydrogen-poor (Type~Ib/c) supernovae whose spectra were dominated by relatively narrow ($\sim 2000$~km~s$^{-1}$) helium features. A few objects with roughly equal-strength narrow hydrogen and helium features have been classified as transitional SNe~Ibn/IIn \citep{Pastorello2}. In all these cases, interaction also adds extra luminosity to the light curves of these events by converting kinetic energy of the ejected material into radiation, placing interacting supernovae near the boundary of normal supernovae and the so-called superluminous supernovae \citep{Gal-Yam2012}.

\setlength{\tabcolsep}{2pt}
\begin{deluxetable*}{lcccclcll}
\tablecaption{Discovery Information}
\tablehead{\colhead{Name} & \colhead{R.A.} & \colhead{Decl.} & \colhead{Redshift} & \colhead{Off-} & \colhead{Disc.} & \colhead{Disc.} & \colhead{Subclass\tablenotemark{c}} & \colhead{Discovery/Classification} \\ & \colhead{(J2000)} & \colhead{(J2000)} & & \colhead{set\tablenotemark{a}} & \colhead{Date\tablenotemark{b}} & \colhead{Mag.} & & \colhead{Reference\tablenotemark{d}}}
\startdata
PTF11rfh         & \ra{22}{35}{49}{58} & \dec{-00}{22}{26}{1} & 0.060 & \phn 0.9 & Dec 08 & 19.8 & emission & this work \\
PTF12ldy         & \ra{23}{45}{11}{04} & \dec{+23}{09}{18}{8} & 0.106 & \phn 3.2 & Nov 12 & 19.7 & P Cygni & this work \\
iPTF14aki\tablenotemark{e} & \ra{14}{20}{41}{73} & \dec{+03}{16}{01}{3} & 0.064 & \phn 2.3 & Apr 10 & 19.9 & emission & \cite{ATel6091}/this work \\
SN 2015U\tablenotemark{f} & \ra{07}{28}{53}{87} & \dec{+33}{49}{10}{6} & 0.014 & \phn 6.1 & Feb 13 & 16.4 & P Cygni & \cite{CBET2015U}/\cite{ATel7105} \\
iPTF15ul         & \ra{12}{14}{03}{76} & \dec{+47}{54}{03}{9} & 0.066 & \phn 0.4 & Mar 10 & 18.8 & ? & this work \\
SN 2015G         & \ra{20}{37}{25}{58} & \dec{+66}{07}{11}{5} & 0.005 &    86.7 & Mar 23 & 15.5 & P Cygni? & \cite{CBET2015G}/\cite{ATel7298} \\
iPTF15akq        & \ra{15}{03}{51}{29} & \dec{+33}{33}{18}{1} & 0.109 & \phn 2.6 & Apr 17 & 20.5 & P Cyg.+H & this work \\
\hline \noalign{\vskip 1pt}
SN 1999cq        & \ra{18}{32}{07}{10} & \dec{+37}{36}{44}{3} & 0.026 & \phn 5.3 & Jun 25 & 15.8 & P Cygni? & \cite{IAUC1999cq} \\
SN 2000er        & \ra{02}{24}{32}{54} & \dec{-58}{26}{18}{0} & 0.031 &    32.1 & Nov 23 & 15.1 & P Cygni & \cite{IAUC2000er} \\
SN 2002ao        & \ra{14}{29}{35}{74} & \dec{-00}{00}{55}{8} & 0.005 &    19.9 & Jan 25 & 14.3 & P Cygni? & \cite{IAUC2002ao} \\
SN 2005la        & \ra{12}{52}{15}{68} & \dec{+27}{31}{52}{5} & 0.018 & \phn 9.5 & Nov 30 & 17.6 & P Cyg.+H & \cite{CBET2005la} \\
SN 2006jc        & \ra{09}{17}{20}{78} & \dec{+41}{54}{32}{7} & 0.006 &    15.5 & Oct 09 & 13.8 & emission & \cite{CBET2006jc} \\
SN 2010al        & \ra{08}{14}{15}{91} & \dec{+18}{26}{18}{2} & 0.017 &    11.5 & Mar 13 & 17.8 & P Cygni & \cite{CBET2010al} \\
SN 2011hw        & \ra{22}{26}{14}{54} & \dec{+34}{12}{59}{1} & 0.023 & \phn 8.0 & Nov 18 & 15.7 & emis.+H & \cite{CBET2011hw} \\
PS1-12sk         & \ra{08}{44}{54}{86} & \dec{+42}{58}{16}{9} & 0.054 &    27.0 & Mar 11 & 18.7 & emission & \cite{Sanders2013} \\
LSQ12btw         & \ra{10}{10}{28}{82} & \dec{+05}{32}{12}{5} & 0.058 & \phn 1.9 & Apr 09 & 19.1 & emission & \cite{ATel4037} \\
OGLE12-006       & \ra{03}{33}{34}{79} & \dec{-74}{23}{40}{1} & 0.057 & \nodata & Oct 07\tablenotemark{g} & 19.0\tablenotemark{g} & P Cygni? & \cite{ATel4495}/\cite{ATel4734} \\
iPTF13beo        & \ra{16}{12}{26}{63} & \dec{+14}{19}{18}{0} & 0.091 & \phn 1.5 & May 19 & 20.9 & emission & \cite{Gorbikov2014} \\
LSQ13ccw         & \ra{21}{35}{51}{64} & \dec{-18}{32}{52}{0} & 0.060 &    14.7 & Sep 04 & 19.1 & emission & \cite{ATel5380} \\
SN 2014av        & \ra{09}{00}{20}{02} & \dec{+52}{29}{28}{0} & 0.030 &    11.1 & Apr 19 & 16.2 & ? & \cite{CBET2014av} \\
SN 2014bk        & \ra{13}{54}{02}{42} & \dec{+20}{00}{24}{3} & 0.070 & \phn 0.3 & May 28 & 17.9 & ? & \cite{CBET2014bk} \\
ASASSN-15ed\tablenotemark{h} & \ra{16}{48}{25}{16} & \dec{+50}{59}{30}{7} & 0.049 & \phn 5.5 & Mar 01 & 17.1 & P Cygni & \cite{ATel7163}/\cite{ATel7219} \\
\hline \noalign{\vskip 1pt}
ASASSN-14dd      & \ra{07}{45}{14}{38} & \dec{-71}{24}{14}{0} & 0.018 &    25.2 & Jun 24 & 15.6 & \nodata & \cite{ATel6269}/\cite{ATel6293} \\
PS15dpn\tablenotemark{i} & \ra{02}{32}{59}{75} & \dec{+18}{38}{07}{0} & 0.175 & \nodata & Dec 29 & 20.7 & \nodata & \cite{GCN18786}/\cite{GCN18811} \\
SN 2016Q         & \ra{08}{10}{19}{86} & \dec{+19}{26}{48}{2} & 0.103 &    28.8 & Jan 07 & 20.4 & \nodata & \cite{ATel8546}
\enddata
\tablecomments{The table is separated into three groups: supernovae with new data reported here (top), supernovae from the literature used in the sample analysis below (middle), and supernovae classified as Type~Ibn but not included in our analysis (bottom).
\tablenotetext{a}{Radial offset from the center of the host galaxy, in arcseconds.}
\tablenotetext{b}{Month and UTC day the supernova was discovered. The year is included in the supernova name.}
\tablenotetext{c}{Spectral subtype of the supernova (see Section~\ref{sec:qual_class}). Supernovae labeled ``P Cygni?'' do not have early spectra available, but their late spectra resemble late spectra of SN~2010al. ``+H'' indicates so-called ``transitional'' Ibn/IIn supernovae that exhibit equal-strength helium and hydrogen emission lines.}
\tablenotetext{d}{Discovery and/or classification notices (ATel, CBET, or IAUC), if any. Refereed publications are listed for PTF and PanSTARRS1 supernovae, which are not publicly announced.}
\tablenotetext{e}{iPTF14aki was independently discovered by the Catalina Real-Time Transient Survey and given the name CSS140421:142042+031602. \cite{ATel6091} released a public classification. iPTF's discovery and classification occurred earlier but were not released publicly.}
\tablenotetext{f}{Previously published under its temporary designation, PSN~J07285387+3349106.}
\tablenotetext{g}{The discovery date is not available, so the first detection is given. The discovery notice was sent on October 17.}
\tablenotetext{h}{Also known as PS15nk.}
\tablenotetext{i}{Also known as iPTF15fgl. Discovered during electromagnetic follow-up of the gravitational wave event GW151226 and only recently disembargoed \citep{Smartt2016a}.}
}
\label{tab:disc}
\end{deluxetable*}
\setlength{\tabcolsep}{6pt}

Of these classes, only Type~IIn supernovae have had direct progenitor detections. The first such case was for SN~2005gl, which \cite{Gal-Yam2009} reported to be an explosion of a luminous blue variable (LBV) star, a class of objects known to undergo episodic mass loss. LBV progenitors were later suggested for Type~IIn SNe~2009ip and 2010jl as well \citep[O.~D.~Fox et al.\ 2016, in preparation]{Smith2010,Smith2010a,Foley2011}, although there is still no consensus that the former was a genuine supernova. Detections of pre-explosion variability at the locations of other Type~IIn supernovae, indicative of eruptions or outbursts \citep[e.g.,][]{Mauerhan2013,Ofek2013,Ofek2014}, further support this scenario.

SNe~Ibn, on the other hand, have only indirect progenitor evidence. One popular narrative, first put forward by \cite{Pastorello2007}, is that these supernovae are the core-collapse explosions of very massive Wolf--Rayet (WR) stars embedded in helium-rich CSM. WR atmospheres are nearly hydrogen-free, and their strong winds could eject material at velocities consistent with the widths of the relatively narrow lines observed in SNe~Ibn \citep[1000--2000~km~s$^{-1}$; see review by][]{Crowther2007}. Alternatively, Type~Ibn progenitors could be members of massive binaries, in which CSM is produced by stripping material from the envelopes of one or both stars \citep{Foley2007}. We discuss possible progenitor channels for SNe~Ibn in more detail in Section~\ref{sec:discuss}.

The best-studied SN~Ibn is SN~2006jc \citep{Foley2007,Pastorello2007,Pastorello1,Immler2008,Smith2008}, discovered by \cite{CBET2006jc} in the relatively nearby ($z=0.006$) galaxy UGC~4904. \cite{CBET2006jc} noted the detection of an optical transient at the coordinates of the supernova two years before the final explosion, as seen in some Type~IIn supernovae. This was seen as evidence for unstable mass loss from a WR progenitor, although no such outbursts have ever been seen from a known WR star \citep{Pastorello2007}. Unfortunately, the light curve of SN~2006jc was already declining at the time of discovery, so we do not have data from immediately after explosion.

However, not all evidence points to a massive progenitor. PS1-12sk, for example, was an SN~Ibn with spectra nearly identical to those of SN~2006jc that exploded in an apparently non-star-forming host \citep{Sanders2013}. Given the extremely low fraction of core-collapse supernovae that occurs in elliptical galaxies \citep[$\lesssim 0.2\%$;][]{Hakobyan2012}, observing one out of fewer than two dozen SNe~Ibn with an elliptical host is unlikely by chance, calling into question whether these are truly explosions of massive stars. \cite{Sanders2013}, for example, suggest several degenerate progenitor scenarios for SNe~Ibn.

One way to distinguish between the various progenitor scenarios is to obtain a large sample of well-observed events, both photometrically and spectroscopically. Unfortunately, many events, especially those discovered prior to 2010, do not have well-observed light curves and spectral series. In the most comprehensive study to date, \cite{Pastorello9} point out that the Type~Ibn class, defined by their spectra, include some light curve outliers. OGLE-2012-SN-006 (hereinafter OGLE12-006), for example, has a very slow decline, whereas LSQ13ccw has a very fast decline and an unusually faint peak magnitude. SNe~2005la and 2011hw, both transitional Type~Ibn/IIn events, as well as iPTF13beo, have double-peaked light curves. SN~2010al, still the earliest observed SN~Ibn relative to maximum light, had an unprecedented premaximum spectrum and an unusually slow rise. However, we show below that Type~Ibn light curves, when taken as a whole, are much more homogeneous and faster evolving than their Type~IIn counterparts.

Here we present photometry and spectroscopy for six new SNe~Ibn: PTF11rfh, PTF12ldy, iPTF14aki, iPTF15ul, SN~2015G, and iPTF15akq. With the addition of these objects, the sample size of SNe~Ibn with published data increases to 22. We also present new data for SN~2015U, previously discussed by \cite{Tsvetkov2015}, \cite{Pastorello8}, and \cite{Shivvers2016}. We then examine light curves and spectra of objects in this sample to infer the properties of the class as a whole. Finally, we discuss the implications of these findings for possible progenitors.

\section{Observations}
\begin{deluxetable}{lclcl}
\tablecaption{Optical and Ultraviolet Photometry}
\tablehead{\colhead{Supernova} & \colhead{MJD} & \colhead{Filter} & \colhead{Magnitude} & \colhead{Source}}
\startdata
PTF11rfh & 55902.109 & R & $>20.73$ & P48 \\
PTF11rfh & 55903.121 & R & $19.91\pm0.10$ & P48 \\
PTF11rfh & 55906.204 & r & $18.70\pm0.04$ & P60 \\
PTF11rfh & 55919.107 & B & $18.59\pm0.09$ & KAIT \\
PTF11rfh & 55919.112 & V & $18.13\pm0.08$ & KAIT \\
PTF11rfh & 55919.117 & R & $18.31\pm0.07$ & KAIT \\
PTF11rfh & 55919.122 & I & $17.70\pm0.07$ & KAIT \\
PTF11rfh & 55928.122 & i & $20.43\pm0.18$ & P60 \\
PTF11rfh & 55928.123 & r & $20.37\pm0.20$ & P60 \\
PTF11rfh & 55928.127 & g & $19.52\pm0.07$ & P60 \\
PTF11rfh & 55930.119 & i & $20.36\pm0.19$ & P60 \\
PTF11rfh & 55930.120 & r & $20.20\pm0.14$ & P60 \\
PTF11rfh & 55941.095 & r & $>20.99$ & P60 \\
PTF11rfh & 55941.098 & g & $20.60\pm0.22$ & P60 \\
PTF11rfh & 55947.088 & i & $>20.53$ & P60 \\
PTF11rfh & 55947.091 & r & $>20.87$ & P60 \\
PTF11rfh & 55947.094 & g & $20.49\pm0.25$ & P60
\enddata
\tablecomments{(This table is available in its entirety in machine-readable form.)
\label{tab:phot_11rfh}
\label{tab:phot_12ldy}
\label{tab:phot_14aki}
\label{tab:phot_15U}
\label{tab:phot_15G}}
\end{deluxetable}

Throughout this paper, times are given as dates in coordinated universal time (UTC) or as modified Julian dates (MJD). ``Phase'' denotes days since maximum light, in the $r$- or $R$-band where available, in the rest frame of the supernova.

PTF11rfh and PTF12ldy were discovered by the Palomar Transient Factory \citep[PTF;][]{P48,PTF} on 2011 December 8 and 2012 November 12, respectively. iPTF14aki, iPTF15ul, and iPTF15akq were discovered by the Intermediate Palomar Transient Factory \citep[iPTF;][]{iPTF} on 2014 April 10, 2015 March 10, and 2015 April 17, respectively. iPTF14aki was independently discovered by the Catalina Real-Time Transient Survey \citep{CRTS} on 2014 April 21, given the name CSS140421:142042+031602, and classified by the Public ESO Spectroscopic Survey of Transient Objects \citep[PESSTO;][]{ATel6091,PESSTO}. SN~2015U was discovered by the Lick Observatory Supernova Search \citep{LOSS,CBET2015U} on 2015 February 13 and classified by \cite{ATel6091}. SN~2015G was discovered by Kunihiro Shima on 2015 March 23 \citep{CBET2015G} and classified by \cite{ATel7298}. Discovery details are given in Table~\ref{tab:disc}.

Multiband optical photometry was obtained using the Catalina Sky Survey \citep[CSS;][]{CRTS} telescope at Steward Observatory (AZ, USA), the 0.76~m Katzman Automatic Imaging Telescope \citep[KAIT;][]{LOSS} at Lick Observatory (CA, USA), the 1~m and 2~m telescopes of the Las Cumbres Observatory Global Telescope Network \citep[LCOGT;][]{LCOGT}, the Nordic Optical Telescope (NOT) at the Observatorio del Roque de los Muchachos (ORM; Canary Islands, Spain), the New Technology Telescope (NTT) at La Silla Observatory (Coquimbo Region, Chile), the 48~inch (1.2~m) Samuel Oschin Telescope (P48) and the 60~inch (1.5~m) telescope \citep[P60;][]{P60} at Palomar Observatory (California, USA). Ultraviolet photometry was obtained with the Ultra-Violet/Optical Telescope \citep[UVOT;][]{UVOT} on board the {\it Swift} satellite.

\begin{figure*}[p]
\centering
\includegraphics[height=0.9\textheight]{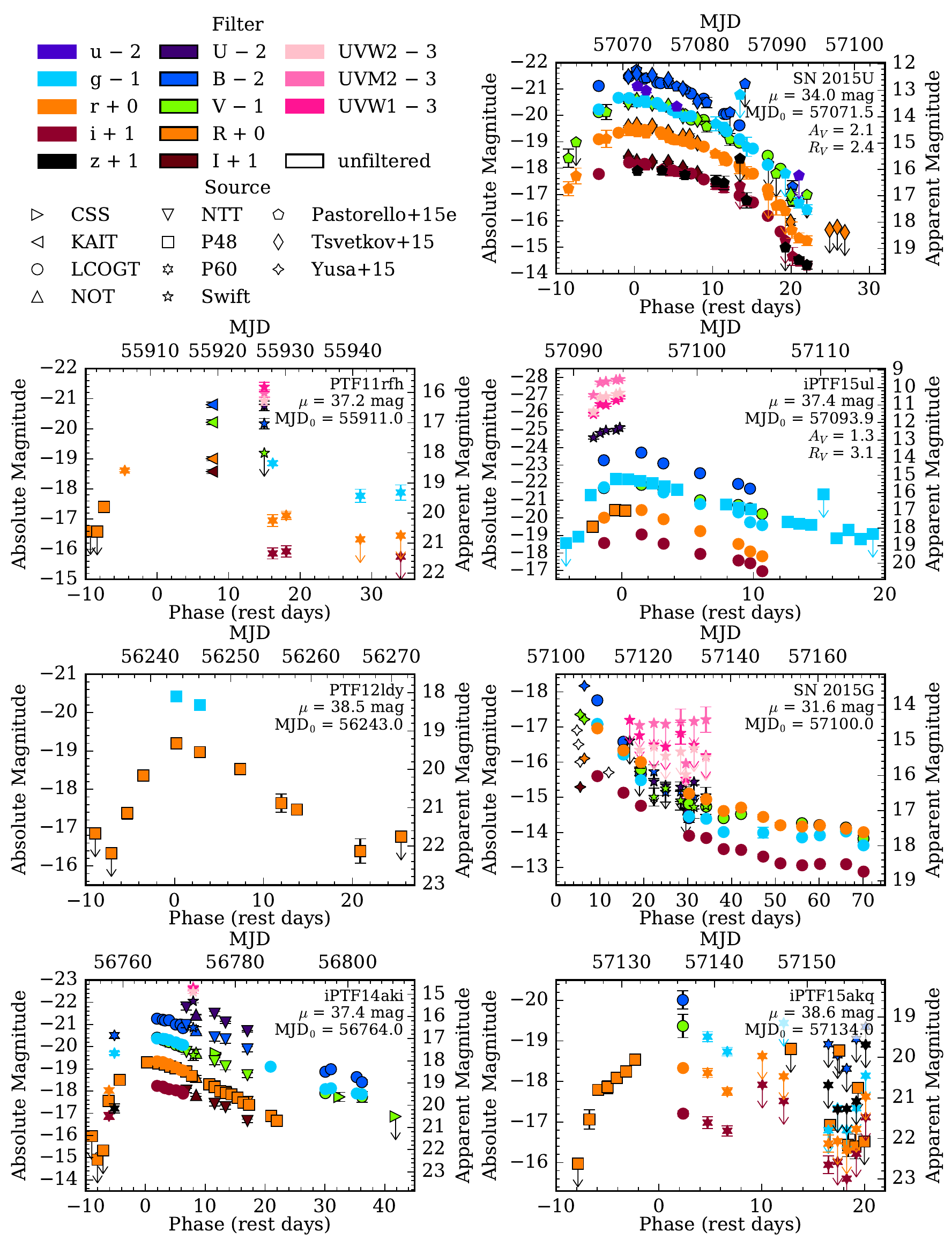}
\caption{Light curves of PTF11rfh, PTF12ldy, iPTF14aki, SN 2015U, iPTF15ul, SN 2015G, and iPTF15akq. Marker shape denotes the source of the photometry (acronyms are defined in the text), and marker color denotes the filter band. Downward-pointing arrows mark $3\sigma$ nondetection upper limits. The right axis for each frame gives the \emph{extinction-corrected} (including host extinction, if $A_V$ and $R_V$ are given) apparent magnitudes in the UVOT system for {\it Swift} points, in the Vega system for other $UBVRI$ points, and in the AB system for $ugriz$ points. The left axis gives the absolute magnitude assuming the distance modulus ($\mu$) given at the top right of each panel. The bottom axis shows rest-frame days from the estimated peak date (MJD$_0$). For SN 2015U, the photometry published by \cite{Tsvetkov2015} and \cite{Pastorello8} is plotted for comparison. For SN 2015G, photometry from the discovery CBET \citep{CBET2015G} is plotted for comparison.}
\label{fig:phot}
\end{figure*}

Each photometry point is listed along with its source in Table~\ref{tab:phot_11rfh} and plotted in Figure~\ref{fig:phot}. All upper limits reported in the tables and figures are $3\sigma$ nondetections. {\it Swift} magnitudes are in the UVOT system \citep{Poole2007}. All other magnitudes are in the Vega system for $UBVRI$ points and in the AB system for $ugriz$ points. (Note that P48 uses a Mould $R$ filter.) See Appendix~\ref{sec:phot_reduce} for details on the reduction process for photometry.

\defcitealias{Planck2015}{Planck Collaboration (2015)}
\defcitealias{extinctionlaw}{CCM89}

After calibration, apparent magnitudes were corrected for Milky Way extinction according to the dust maps of \cite{dustmaps}, obtained via the NASA/IPAC Extragalactic Database (NED). SN~2015U and iPTF15ul were also corrected for their high host-galaxy extinction using the extinction law of \cite{extinctionlaw} with the best-fit parameters derived in Section~\ref{sec:extinction}. Absolute magnitudes were obtained using the distance moduli calculated from host-galaxy redshifts, if available, or redshifts obtained from the supernova spectra, and the cosmological parameters presented by the \citetalias{Planck2015}. None of the mean redshift-independent distance moduli listed on NED for these objects are more than $3\sigma$ away from the values used here.

For iPTF14aki, two epochs of near-infrared photometry were obtained with the Near Infrared Camera Spectrometer (NICS) on the Telescopio Nazionale Galileo (TNG) at ORM. Magnitudes were measured using aperture photometry and calibrated to nearby stars in the 2MASS catalog \citep{2MASS}. We only detect the supernova at $>3\sigma$ significance in the $J$-band: MJD 56779.9, $J = 18.4 \pm 0.2$~mag, $H > 18.9$~mag, $K_s > 18.9$~mag; MJD 56794.1, $J = 19.4 \pm 0.2$~mag, $H > 18.7$~mag, $K_s > 18.0$~mag.

We also observed PTF11rfh, along with the necessary calibrators, with the Karl G.\ Jansky Very Large Array \citep[VLA;][]{Perley2009} under our Target of Opportunity program (VLA/11A-227; PI: A.~Corsi) on 2011 December 31.8 with the VLA in its D configuration. VLA data were reduced and imaged using the Common Astronomy Software Applications (CASA) package \citep{CASA}. Our observation yielded a nondetection, with a $3\sigma$ upper limit of $\lesssim 81$~\textmu Jy at 6.4~GHz.

Optical spectra were obtained using the Andalucia Faint Object Spectrograph and Camera (ALFOSC) on NOT; the Double Spectrograph \citep[DBSP;][]{DBSP} on the 200~inch (5.1~m) Hale Telescope (P200) at Palomar Observatory; the Double Imaging Spectrograph (DIS) on the Apache Research Consortium (ARC) 3.5~m telescope at Apache Point Observatory (NM, USA); the Device Optimized for the Low Resolution (DOLORES) on TNG; the ESO Faint Object Spectrograph and Camera (EFOSC2) on NTT; the FLOYDS robotic spectrographs on the LCOGT 2~m telescopes at Haleakal\={a} Observatory (HI, USA) and Siding Spring Observatory (New South Wales, Australia); the Kast Double Spectrograph on the 3~m C.~Donald Shane Telescope at Lick Observatory; and the Low-Resolution Imaging Spectrometer \citep[LRIS;][]{LRIS} on the Keck~I telescope, the Deep Imaging Multi-Object Spectrograph \citep[DEIMOS;][]{DEIMOS} on the Keck~II telescope, and the Gemini Multi-Object Spectrograph \citep[GMOS;][]{GMOS} on the Frederick C.\ Gillett Gemini North telescope, all on Mauna Kea (HI, USA). The near-infrared spectrum of SN~2015U was obtained with the SpeX instrument \citep{SpeX} at the NASA Infrared Telescope Facility (IRTF), also on Mauna Kea.

Each spectrum is logged in Table~\ref{tab:spec} and plotted in Figures~\ref{fig:06jclike_spec}--\ref{fig:15ul_spec}. They are not corrected for reddening. See Appendix~\ref{sec:spec_reduce} for details on the reduction process for spectroscopy. All spectra in Table~\ref{tab:spec} have been submitted to the Weizmann Interactive Supernova Data Repository \citep[WISeREP;][]{WISeREP}.

\setlength{\tabcolsep}{4pt}
\begin{deluxetable}{lcllcc}
\tablecaption{Log of Spectroscopic Observations}
\tablehead{\colhead{Supernova} & \colhead{MJD} & \colhead{Telescope} & \colhead{Instrument} & \multicolumn{2}{c}{Days From\tablenotemark{a}} \\ \cline{5-6} \\[-6pt] & & & & \colhead{Peak} & \colhead{Expl.}}
\startdata
PTF11rfh  & 55912.9   & TNG      & DOLORES & $+2$ & 10 \\
          & 55916.1   & P200     & DBSP    & $+5$ & 13   \\
          & 55926.2   & Keck I   & LRIS    & $+14$ & 22  \\
          & 55929.\phn & Lick 3 m & Kast    & $+17$ & 25  \\
\hline \noalign{\smallskip}
PTF12ldy  & 56244.\phn & ARC 3.5 m& DIS     & $+1$ & \phn7   \\
          & 56246.\phn & Keck II  & DEIMOS  & $+3$ & \phn9   \\
          & 56270.\phn & Keck II  & DEIMOS  & $+24$ & 31  \\
          & 56279.\phn & Keck I   & LRIS    & $+32$ & 39  \\
\hfill \tablenotemark{b} & 56300.\phn & Keck II  & DEIMOS  & $+51$ & 58 \\
\hline \noalign{\smallskip}
iPTF14aki & 56765.1   & NOT      & ALFOSC  & $+1$ & \phn8   \\
          & 56769.6   & LCOGT    & FLOYDS  & $+5$ & 12   \\
          & 56770.2   & NTT      & EFOSC2  & $+6$ & 13   \\
          & 56771.2   & NTT      & EFOSC2  & $+7$ & 14   \\
          & 56773.0   & NOT      & ALFOSC  & $+8$ & 16   \\
\hfill \tablenotemark{c}  & 56774.\phn & Keck II  & DEIMOS  & $+9$ & 17 \\
          & 56774.1   & TNG      & DOLORES & $+9$ & 17   \\
          & 56776.2   & NTT      & EFOSC2  & $+11$ & 18  \\
          & 56778.1   & NTT      & EFOSC2  & $+13$ & 20  \\
          & 56782.1   & NTT      & EFOSC2  & $+17$ & 24  \\
          & 56784.1   & NTT      & EFOSC2  & $+19$ &  26 \\
          & 56786.4   & P200     & DBSP    & $+21$ & 28  \\
          & 56805.4   & Keck I   & LRIS    & $+39$ & 46  \\
\hline \noalign{\smallskip}
SN 2015U  & 57070.4   & LCOGT    & FLOYDS  & $-1$ & \phn8   \\
          & 57076.2   & IRTF     & SpeX    & $+5$ & 15   \\
          & 57076.5   & LCOGT    & FLOYDS  & $+5$ & 15   \\
\hfill \tablenotemark{b} & 57096.3   & LCOGT    & FLOYDS  & $+24$ & 34 \\
\hline \noalign{\smallskip}
iPTF15ul  & 57092.\phn & NOT      & ALFOSC & $-2$ & \phn1    \\
          & 57093.\phn & NOT      & ALFOSC & $-1$ & \phn2    \\
          & 57102.\phn & Gemini N & GMOS   & $+8$ & 11    \\
          & 57109.\phn & NOT      & ALFOSC & $+14$ & 17   \\
\hline \noalign{\smallskip}
SN 2015G  & 57122.5   & LCOGT    & FLOYDS  & $+22$ & \nodata   \\
          & 57124.6   & LCOGT    & FLOYDS  & $+24$ & \nodata  \\
          & 57131.6   & LCOGT    & FLOYDS  & $+31$ & \nodata  \\
          & 57144.5   & LCOGT    & FLOYDS  & $+44$ & \nodata   \\
\hline \noalign{\smallskip}
iPTF15akq & 57134.5   & Keck I   & LRIS    & $+0$ & \phn9   \\
          & 57135.6   & Keck I   & LRIS    & $+1$ & 10   \\
          & 57138.\phn & NOT      & ALFOSC  & $+4$ & 12   \\
          & 57149.4   & Gemini N & GMOS    & $+14$ & 22  \\
          & 57158.5   & Keck I   & LRIS    & $+22$ & 30 \enddata
\tablecomments{\tablenotetext{a}{Rest-frame days from peak and from explosion, if known.}
\tablenotetext{b}{Dominated by host galaxy.}
\tablenotetext{c}{Not shown in Figure~\ref{fig:06jclike_spec} due to low signal-to-noise ratio.}
\label{tab:spec}}
\end{deluxetable}
\setlength{\tabcolsep}{6pt}

\section{Extinction Estimation}\label{sec:extinction}
As first noted by \cite{ATel7105}, SN~2015U suffers from significant host-galaxy extinction. The extinction parameters, parametrized using the extinction law of \citet[hereafter \citetalias{extinctionlaw}]{extinctionlaw} have been estimated previously by at least two groups: \cite{Pastorello8} estimate $E(B-V) = 0.99 \pm 0.48$~mag assuming a fixed $R_V = 3.1$, and \cite{Shivvers2016} estimate $E(B-V) = 0.94_{-0.4}^{+0.1}$~mag for $R_V = 2.1$ by fitting reddened blackbody models to photometry of SN~2015U.

We compare our optical through near-infrared spectrum of SN~2015U at $+5$~days to broadband photometry and spectra of SN~2010al around the same phase. This yields a best-fit extinction law with $E(B-V) = 0.86 \pm 0.10$~mag and $R_V = 2.4 \pm 0.2$ (as parametrized by \citetalias{extinctionlaw}). Figure~\ref{fig:15U_spec} shows a reddened optical--near-infrared spectrum of SN~2010al \citep[from][]{Pastorello4} at $-1$~day to match the observed continuum shape of SN~2015U.

We performed a similar analysis comparing iPTF15ul to SN~2000er. Lacking near-infrared data for iPTF15ul, we do not attempt to fit the total-to-selective extinction ratio, $R_V$. The resulting best-fit extinction parameters are $E(B-V) = 0.41 \pm 0.10$~mag with $R_V = 3.1$. This is illustrated in Figure~\ref{fig:15ul_spec}, where we show a reddened spectrum of SN~2000er \citep[from][]{Pastorello1} together with the observed spectra of iPTF15ul.

This method assumes that SN~2015U and iPTF15ul follow a similar color evolution to SNe~2010al and 2000er, respectively. Under this assumption, our extinction estimate ($A_R = 1.07 \pm 0.26$~mag) makes iPTF15ul the most luminous SN~Ibn studied so far ($M_R = -20.64 \pm 0.29$~mag). However, given the small number of SNe~Ibn with well-sampled multicolor light curves, our result should be treated with some caution.

We assume negligible host-galaxy extinction for the other five supernovae discussed here, as they are either well separated from their host galaxies or their hosts are relatively faint. The continuum shapes of their early spectra support this assumption.

\begin{figure*}
\centering
\includegraphics[height=0.9\textheight]{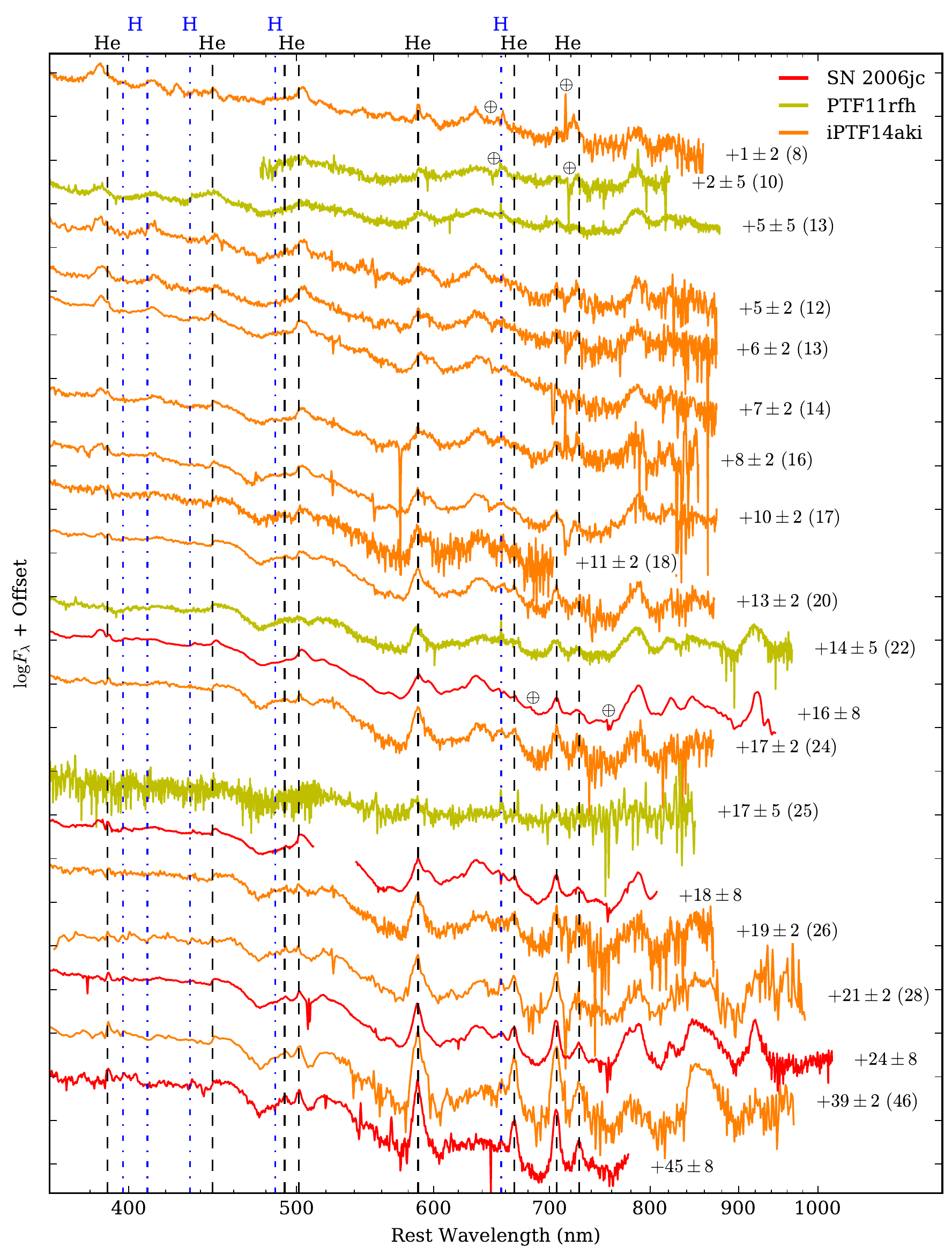}
\caption{Combined spectral series of PTF11rfh and iPTF14aki with representative spectra of SN~2006jc for comparison. Phase is indicated to the right of each spectrum, with rest-frame days from explosion in parentheses, where available. Vertical lines mark emission wavelengths of various elements, listed along the top axis. The Earth symbol ($\oplus$) marks wavelengths affected by telluric absorption once for each supernova. Tick marks on the vertical axes are 0.5~dex apart. Note the similarity between these three objects and the contrast with the early spectra in Figure~\ref{fig:pcygni_spec}.}
\label{fig:06jclike_spec}
\end{figure*}

\begin{figure*}
\centering
\includegraphics[height=0.9\textheight]{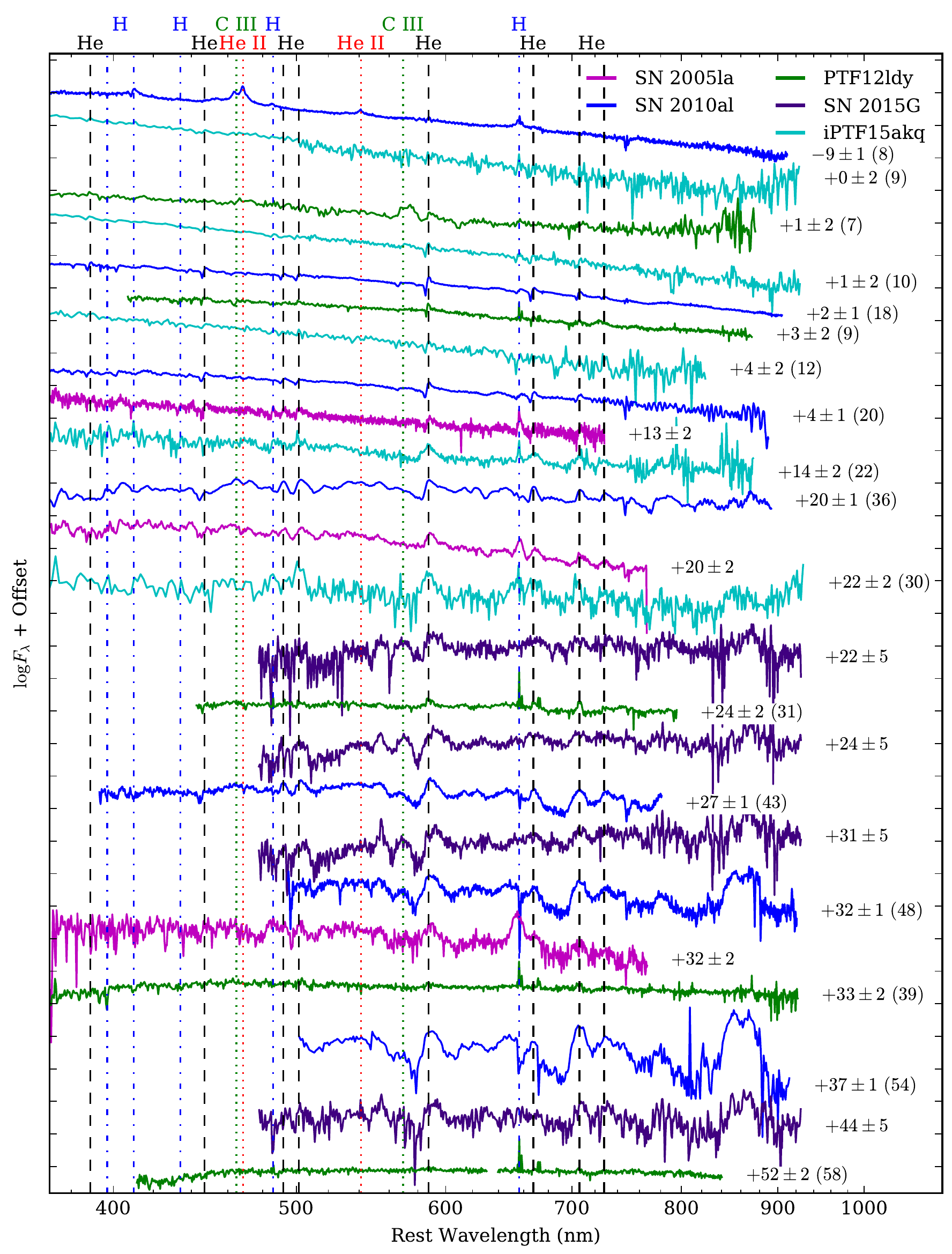}
\caption{Combined spectral series of PTF12ldy, SN~2015G, and iPTF15akq, with representative spectra of SNe~2005la and 2010al for comparison. SN 2005la and iPTF15akq are transitional Ibn/IIn supernovae, meaning their spectra show relatively strong hydrogen features. Phase is indicated to the right of each spectrum, with rest-frame days from explosion in parentheses, where available. Vertical lines mark emission wavelengths of various elements, listed along the top axis. Tick marks on the vertical axes are 0.5~dex apart. Note the similarity between these objects and contrast with the early spectra in Figure~\ref{fig:06jclike_spec}. Also note the detection of C~III in the first spectrum of PTF12ldy.}
\label{fig:pcygni_spec}
\end{figure*}

\begin{figure*}
\includegraphics[width=\textwidth]{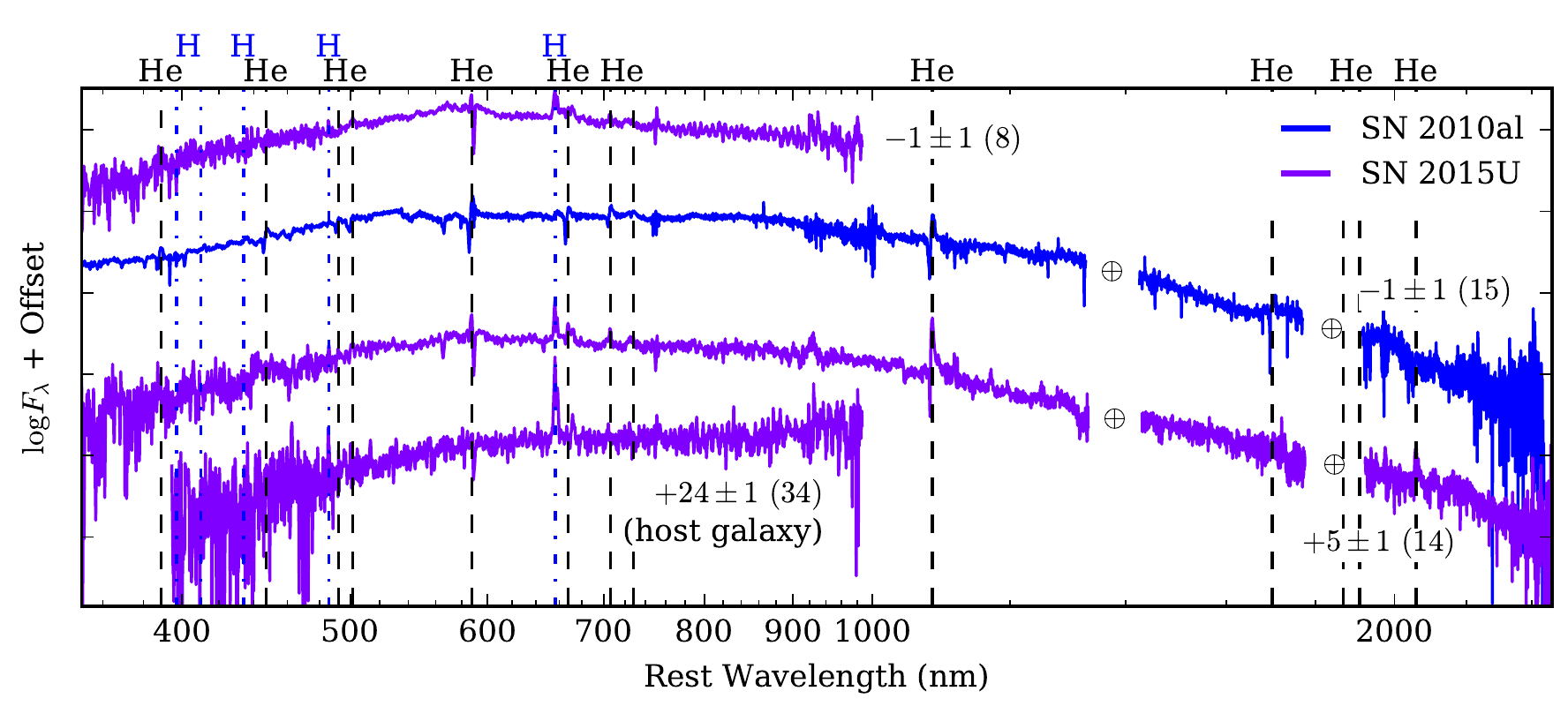}
\caption{Spectra of the highly reddened SN~2015U with an optical through near-infrared spectrum of SN~2010al for comparison. The spectrum of SN~2010al has been reddened using the best-fit parameters for SN~2015U, $A_V = 2.06$~mag and $R_V = 2.4$ (as parametrized by \citetalias{extinctionlaw}). Phase is indicated near the right of each spectrum, with rest-frame days from explosion in parentheses. The spectrum from $+5$ days is a combination of an optical spectrum (range 320--900~nm; resolution 0.2~nm) from FLOYDS and a near-infrared spectrum (range 850--2500~nm; resolution 0.3~nm) from SpeX taken several hours apart. The host-dominated spectrum from $+24$ days is shown in order to identify host lines in the earlier spectra. Vertical lines mark emission wavelengths of various elements, listed along the top axis. The Earth symbol ($\oplus$) marks regions trimmed to remove telluric contamination. Tick marks on the vertical axes are 0.5~dex apart. The P~Cygni profile of the 1083~nm helium line links this to the subclass of objects in Figure~\ref{fig:pcygni_spec}.}
\label{fig:15U_spec}
\end{figure*}

\section{Sample Analysis}

\subsection{Spectroscopy}
In order to compare the spectral evolution of the objects in our sample, we collected all publicly available optical spectra of each object from WISeREP and added them to the spectra listed in Table~\ref{tab:spec}.

\subsubsection{Qualitative Classification}\label{sec:qual_class}
To begin, we examine the spectral series of all the objects in our sample. We find that, at early times ($\lesssim 20$~days after peak), the spectra show some diversity in their helium line profiles. While all Type~Ibn spectra have blue continua relative to most other supernova classes (after considering extinction and host contamination), some show only relatively narrow (1000~km~s$^{-1}$) P~Cygni lines of helium, while others exhibit narrow to intermediate-width (few thousand km~s$^{-1}$) helium emission lines and broad features, sometimes with complex profiles. The latter are most similar to the earliest spectra of the traditional Type~Ibn archetype, SN~2006jc, which unfortunately was not observed around maximum light.

At late times ($\gtrsim 20$ days), the absorption components of the P~Cygni lines become less pronounced and their velocities increase, making the two types of spectra less distinguishable. This means that early spectra of SNe~Ibn may not look like those of the archetype SN~2006jc, similarity to which is often used as a basis for classification, although they will likely resemble it at later times. Early spectral time series will be essential to understanding this spectral diversity.\footnote{This represents a more general problem in supernova classification: classifications are usually announced after the first spectrum is obtained and not revised (at least not publicly) based on subsequent spectra. This can lead to misclassifications for several types of supernovae (e.g., Type~II instead of Type~IIb).}

Until recently, all SNe~Ibn in the literature observed prior to 12 days after peak showed P~Cygni lines in their earliest spectra. However, spectra of PTF11rfh and iPTF14aki, although they resemble other SNe~Ibn at late times, do not show P~Cygni lines at early times. This confirms that Type~Ibn spectra can have at least two morphologies during the first week after maximum light (see Figure~\ref{fig:allspec}).

With our current sparsely sampled, heterogeneous spectral data set, we cannot conclusively say whether these two morphologies arise from two distinct CSM configurations or from a continuum of CSM properties.  A few supernovae do hint at a continuum of line profiles. For example, SNe~2014av and 2014bk both have dominant P~Cygni lines at early times, but these lines are superimposed on broader emission features, not just blue continua. Weaker P~Cygni lines are also superimposed on some emission-dominated spectra, such as those of SN~2006jc, PS1-12sk, and iPTF14aki. Nonetheless, for the purposes of this analysis, we will describe spectra as belonging to one of two different subclasses: the ``P~Cygni'' subclass or the ``emission'' subclass.

Furthermore, we admit that a more physical analysis might group spectra by phase relative to explosion, rather than phase relative to maximum light, but weak constraints on the explosion times for many objects in our sample prevent this. It remains to be seen whether the spectral diversity would persist in such an analysis.

Since iPTF15akq and SNe~2005la and 2011hw exhibit relatively strong hydrogen features throughout their evolution, leading to their transitional classification as Type~Ibn/IIn, it is not clear whether they belong in either of the two proposed classes. Nonetheless, with the exception of the hydrogen lines, their spectra are quite similar to those of other SNe~Ibn, so we group them by line profile as well.

\begin{figure*}
\includegraphics[width=\textwidth]{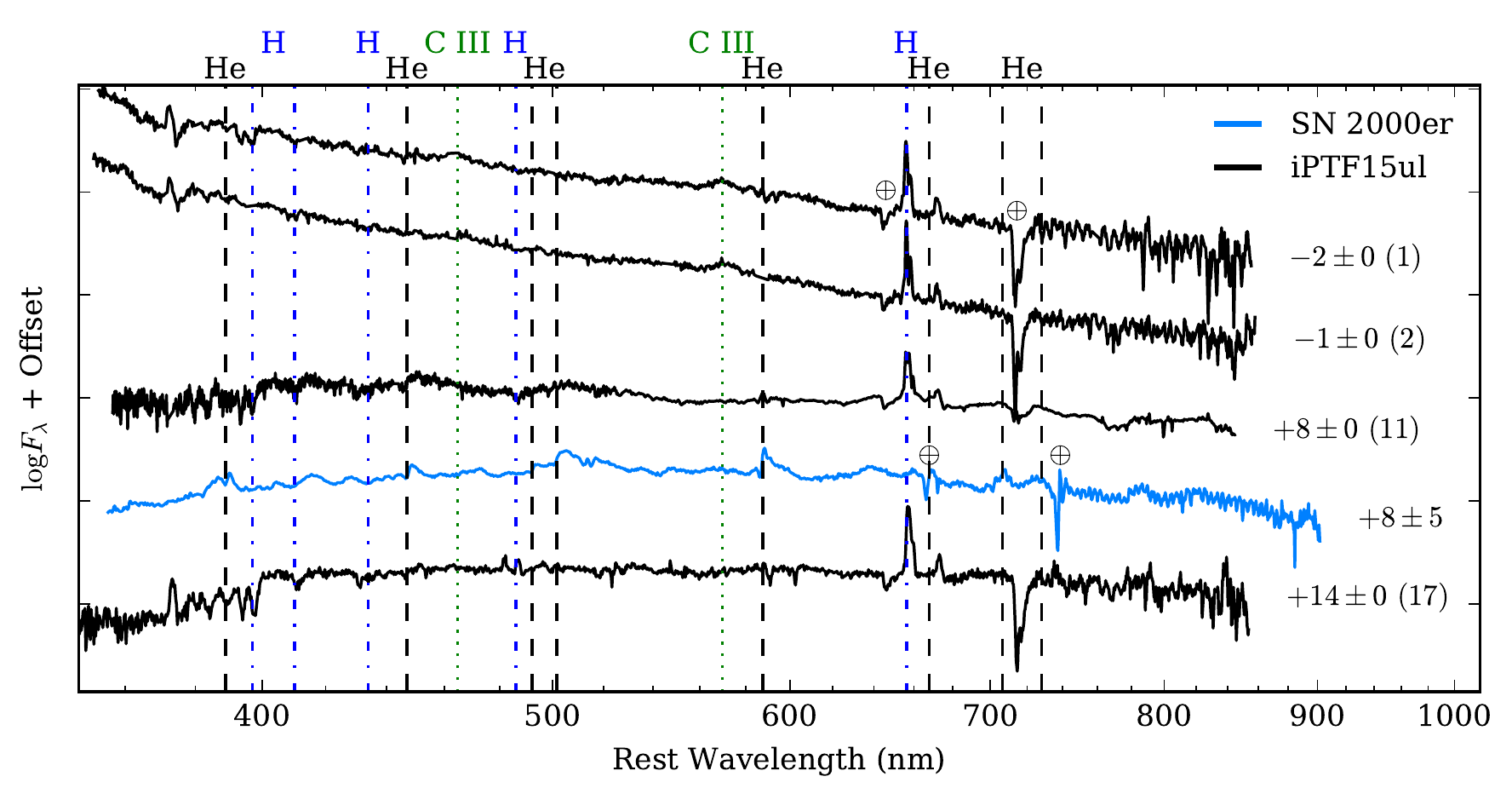}
\caption{Spectra of iPTF15ul with a spectrum of SN~2000er for comparison. The spectrum of SN~2000er has been reddened using the best-fit parameters for iPTF15ul, $A_V = 1.27$~mag and $R_V = 3.1$ (as parametrized by \citetalias{extinctionlaw}). Phase is indicated near the right of each spectrum, with rest-frame days from explosion in parentheses, where available. Vertical lines mark emission wavelengths of various elements, listed along the top axis. The Earth symbol ($\oplus$) marks features affected by telluric absorption. Tick marks on the vertical axes are 0.5~dex apart. Host contamination of iPTF15ul makes it difficult to identify features, but its spectral similarity with SN~2000er allows us to classify it as an SN~Ibn. Also note the detection of C~{\sc iii} in the premaximum spectra of iPTF15ul.}
\label{fig:15ul_spec}
\end{figure*}

\begin{figure}
\includegraphics[width=\columnwidth]{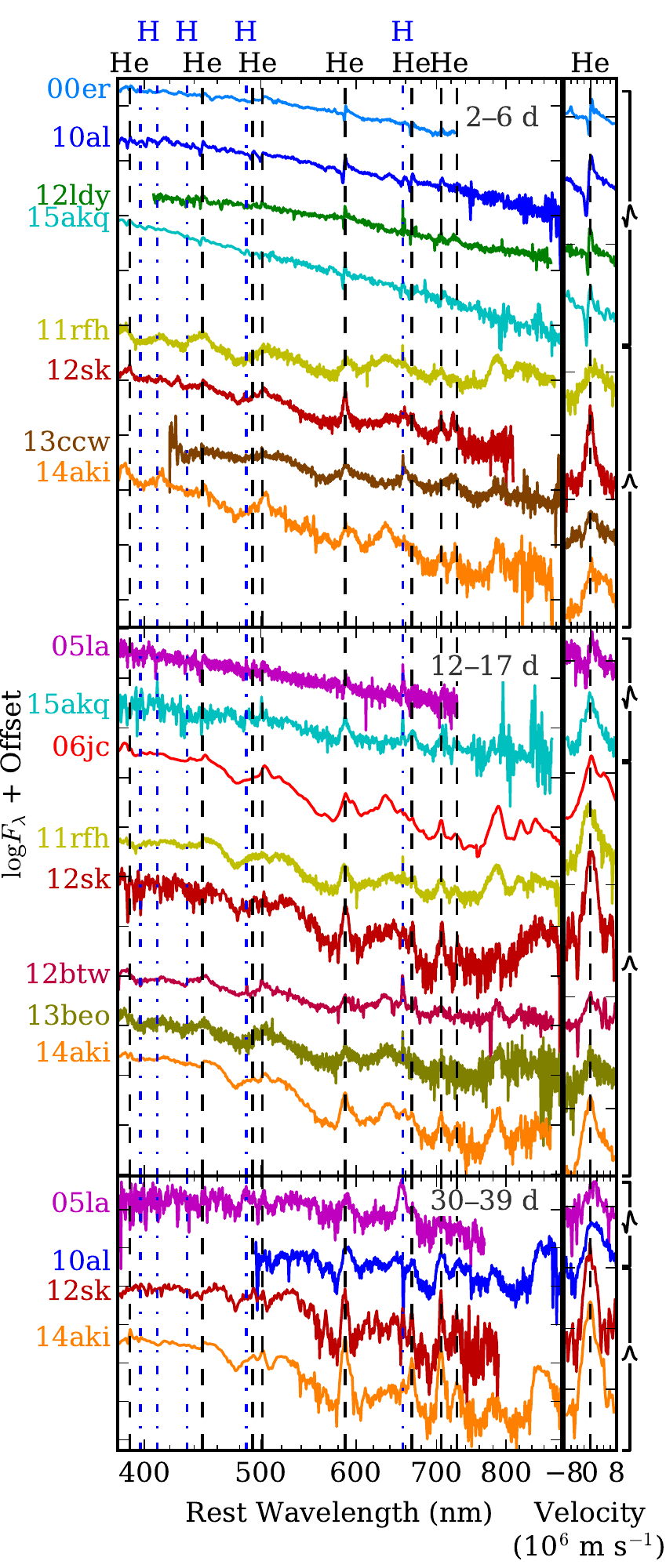}
\caption{Spectra of SNe~Ibn at three phase ranges. Individual supernovae are color-coded throughout this work (see legend in Figure~\ref{fig:vel} for full names): cool colors for the P~Cygni subclass (top of each panel) and warm colors for the emission subclass (bottom of each panel). The column at right shows an expanded view of the helium emission line at 588~nm for each spectrum. Note the relative homogeneity within each subclass and the distinct difference between subclasses at early times ($\lesssim 20$~days after peak). The distinction fades over the course of the first month.}
\label{fig:allspec}
\end{figure}

\begin{figure*}
\centering
\includegraphics[width=\textwidth]{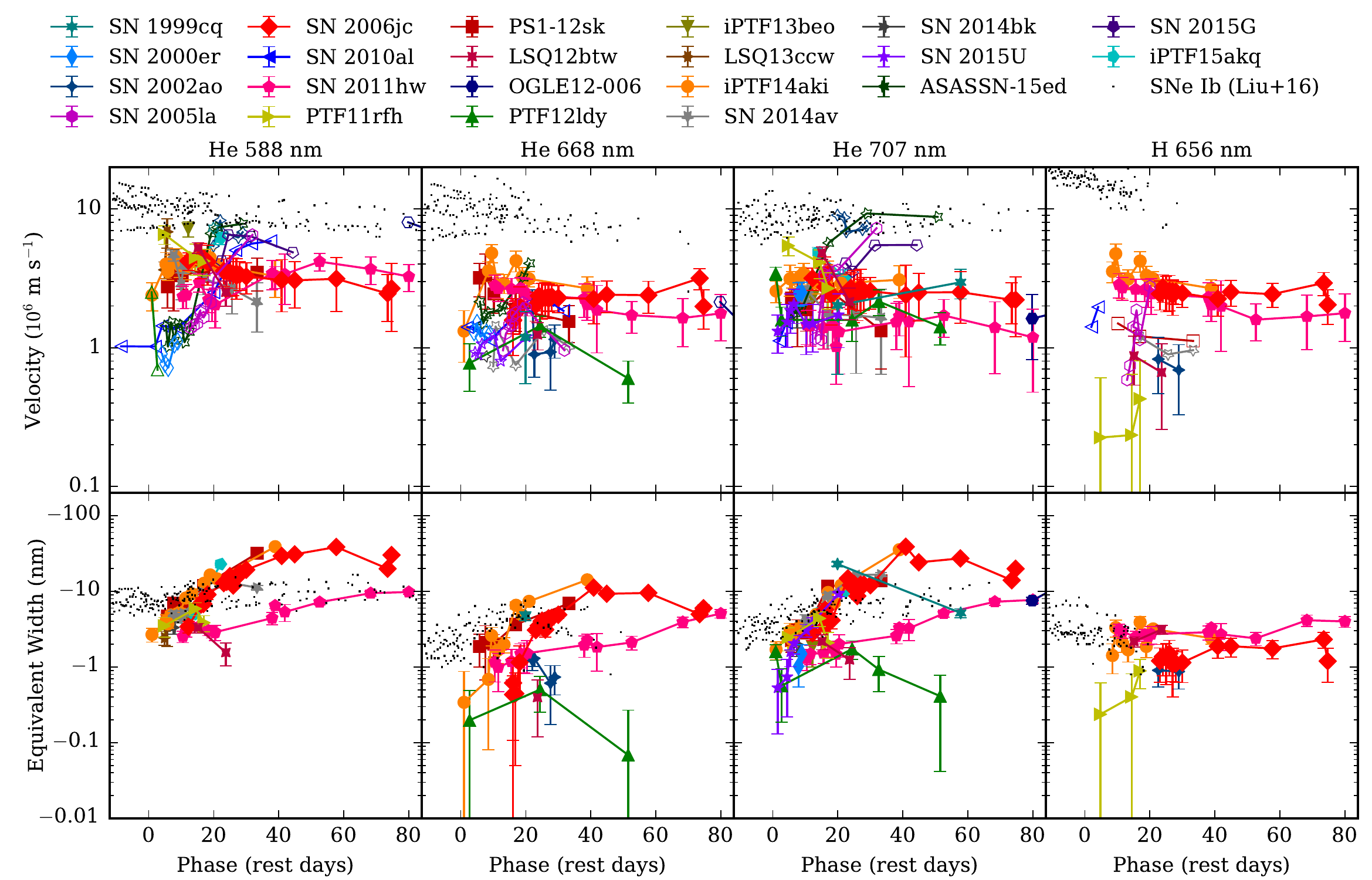}
\caption{Top: velocity evolution of various spectral lines by object. Filled markers indicate pure emission lines, in which case the quoted velocity is the FWHM of the line. Open markers represent lines with P Cygni profiles, in which case the velocity comes from the absorption minimum. The increasing trend for P Cygni velocities likely does not mean the material is accelerating. Rather, at early times, we only see a slow-moving, optically thick CSM shell. As this shell becomes optically thin, we are increasingly able to see the faster-moving supernova ejecta. Otherwise, flat velocity curves simply suggest supernova ejecta in free expansion. Black dots represent absorption velocities of the \cite{Liu2016} sample of SNe~Ib for comparison. Bottom: equivalent width (EW) evolution of various spectral lines by object. Only emission lines are shown. The upward trend in the helium lines suggests that circumstellar helium is being swept up by the ejecta. Black dots represent EWs of helium absorption in SNe~Ib, also from \cite{Liu2016}.}
\label{fig:vel}\label{fig:ew}
\end{figure*}

\subsubsection{Line Analysis}
Of particular interest is the tentative detection of doubly ionized carbon (C~{\sc iii}) at 465.0 and 569.6~nm in the earliest spectra of PTF12ldy (Figure~\ref{fig:pcygni_spec}) and iPTF15ul (Figure~\ref{fig:15ul_spec}). C~{\sc iii} has only appeared once before in a SN~Ibn, in the earliest spectrum ever obtained of this class: that of SN~2010al nine days before maximum light (first spectrum in Figure~\ref{fig:pcygni_spec}). In that case, it appeared alongside other high-ionization species (N~{\sc iii} and He~{\sc ii}), which \cite{Pastorello4} interpreted as a sign of the flash-ionization of the CSM \citep[as in][]{Gal-Yam2014}. To our knowledge, the spectrum of iPTF15ul is the third earliest (relative to peak) Type~Ibn spectrum obtained, suggesting that these lines may still be C~{\sc iii} recombining after shock breakout. We note, however, that the spectrum of SN~2010al showed only the 465.0~nm line, whereas the 569.6~nm lines in our spectra are stronger, possibly calling into question our tentative identification. Si~{\sc ii} could be contributing at 567.0~nm, but we cannot identify any other Si~{\sc ii} features. Although the features are relatively weak, they are notably broader than the narrow P~Cygni lines that develop around maximum light. This may be caused by electron scattering \citep[also as in][]{Gal-Yam2014}.

For each spectrum in our sample, we measure the expansion velocities and equivalent widths (EWs) of three neutral helium (He) lines (588, 668, and 707~nm) and \mbox{hydrogen-$\alpha$} (H$\alpha$, 656~nm), where visible. The results are plotted in Figure~\ref{fig:vel}.

Especially at late times, it is not trivial to distinguish broad P~Cygni features on a blue continuum and broad emission on a flat continuum. We treat more symmetric lines as emission only (filled markers in Figure~\ref{fig:vel}) and more asymmetric lines as P~Cygni (open markers). For simplicity, we do not treat lines with multiple velocity components differently.

We fit emission lines with a Gaussian function on a constant or linear continuum (depending on the object and the line) using a Markov chain Monte Carlo (MCMC) routine based on the {\tt emcee} package \citep{emcee}. The velocity reported is derived from the FWHM intensity of the Gaussian. The EW is the integral of the flux normalized to the local continuum, which may or may not be constant over the width of the line. In cases where a nearby line might interfere with the measurement, we fit and subtract an additional Gaussian to the second line prior to integration.

We fit P~Cygni lines with two Gaussians, one in absorption and one in emission, using the same MCMC routine. The velocity reported is derived from the wavelength difference between absorption and emission. We do not measure EW for P~Cygni lines.

If the line is not clearly identifiable and visible above the noise, or if the width of the line is potentially unresolved by the spectrograph, we do not attempt a measurement. However, since the decision of whether or not to measure is based on a visual inspection, some misidentified lines may be included, and conversely some ambiguous lines may be omitted.

Error bars correspond to standard deviations of the fit-parameter distributions from the MCMC routine. They do not include systematic errors such as choosing the wrong line profile, ignoring multiple velocity components, etc.

Within each of the two spectral subclasses, especially within the emission subclass, there is very little variation in spectral evolution between objects. In most cases we see a general increasing trend in the helium EW, whereas the H$\alpha$ EW, if present at all, stays roughly constant. Line velocities of the emission spectra show little evolution over time. On the other hand, P~Cygni lines visible in early spectra tend to morph into broader emission-dominated lines over the course of the first month. We discuss the implications of these results in Section~\ref{sec:discuss}.

\subsection{Photometry}

\begin{figure*}
\includegraphics[width=\textwidth]{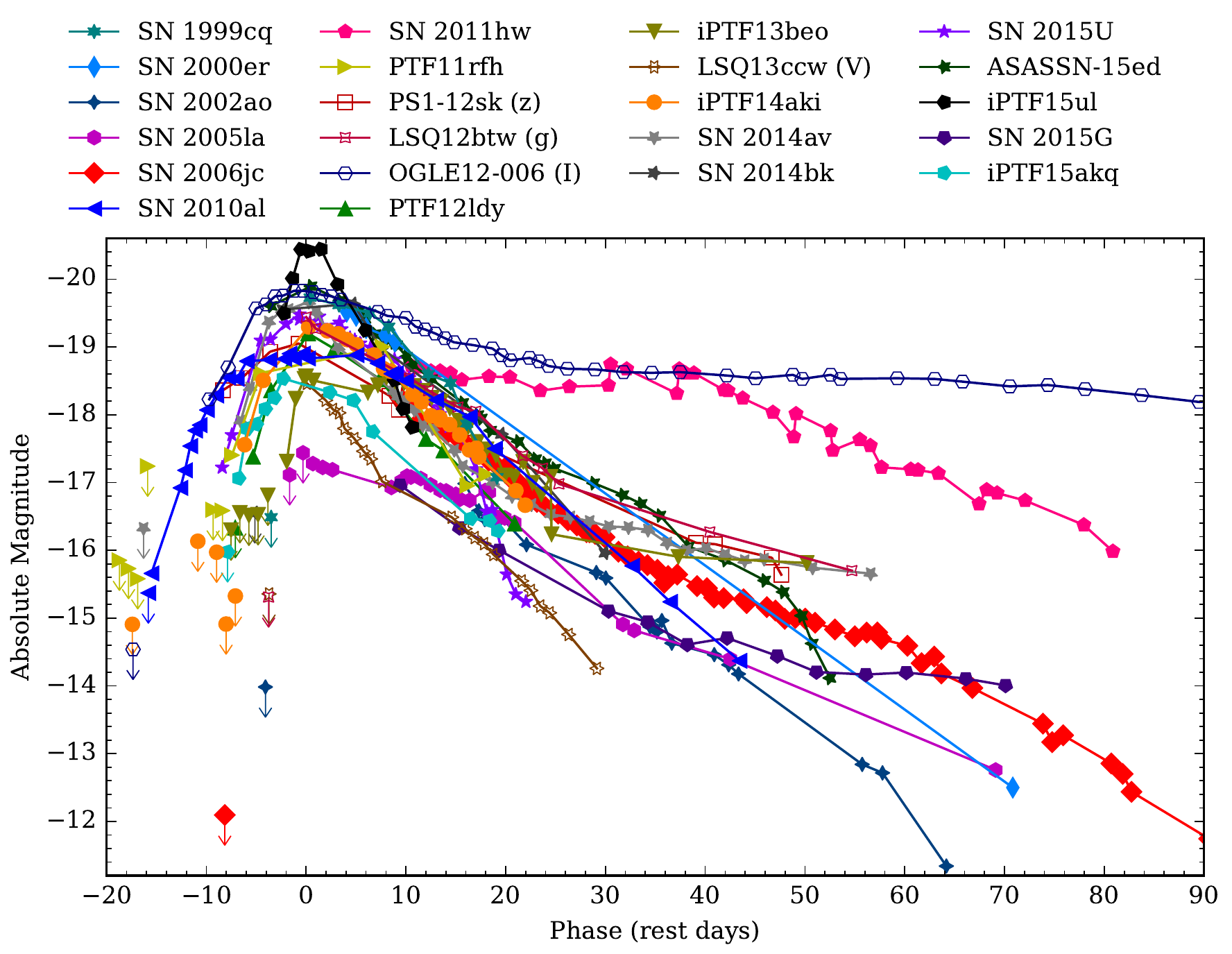}
\caption{Rest-frame light curves used in the sample analysis. The $R$- or $r$- band was chosen unless a different filter had significantly better coverage, in which case an open marker is used and the filter is listed in parentheses. Note the homogeneity in light curve shape, even between subclasses, with a few significant exceptions.}
\label{fig:all_phot}
\end{figure*}

We obtained all published light curves of SNe~Ibn from the literature. The objects we consider, as well as the references used, are listed in Table~\ref{tab:lcparams}. Most, but not all, objects have well-observed $R$- or $r$-band light curves, which we plot in Figure~\ref{fig:all_phot}. Where this was not the case, we used light curves in a better-observed filter and noted this in the ``Filter'' column of Table~\ref{tab:lcparams} and next to the supernova name in Figure~\ref{fig:all_phot}. Although supernovae do not always have the same behavior in all filters, we assume that these differences are small compared to the uncertainties involved in our estimates below. We also do not apply $K$-corrections to the light curves, for the same reason.

Several of the parameters we wish to explore depend on the determination of the time of maximum light of the supernovae, but unfortunately, over half of the objects have few or no points in their light curve around peak brightness. Table~\ref{tab:lcparams} lists our estimates of the peak date, with conservative error bars where the peak is not well constrained. See Appendix~\ref{ap:peaks} for details on how these were chosen. SN~2005la, SN~2011hw, and iPTF13beo have two peaks; in what follows, we consider only the first peak.

\subsubsection{Decline Slope and Peak Magnitude}\label{sec:decline_peakmag}
With an estimate of the epoch of maximum light of each supernova, we can determine the slope of the post-maximum decline. We divide the rest-frame light curve into six 15 day bins, starting with 0--15 days after peak, and five overlapping 30 day bins, starting with 0--30 days after peak. For each of these bins that includes three or more photometry points, we use weighted least-squares regression to determine a best-fit line (magnitude vs.\ phase) to that portion of the light curve. This effectively gives a coarse derivative of the data without fitting a function to the entire light curve. The slopes of these lines are listed in Table~\ref{tab:declineslope} and plotted against the centers of their bins in Figure~\ref{fig:declineslope}.

Most supernovae have several-day errors on the peak date, which in theory affects this calculation. However, since the light curve slopes generally change slowly after maximum light, we neglect this contribution to the error. Errors on the slope are the standard errors of the fit.

SN~2005la, SN~2011hw, and iPTF13beo have a second peak in the light curve during one or more of these bins. We omit any bins where the light curve is not consistent with a monotonically decreasing function. Note that lines in Figure~\ref{fig:declineslope} interpolate over these gaps in coverage, hiding the fact that some slopes become negative (rebrightening) for a short period.

The intercept of the earliest linear fit to the light curves is an estimate of the peak magnitude. In most cases this is the 0--15 day bin. For PTF11rfh and SNe~2002ao, 2014bk, and 2015G, which have fewer than three points in the 0--15 day bin, the intercept from the 0 to 30 day bin is used. The errors on the peak magnitudes include contributions from the estimated error on the peak date, the standard errors on the slope and intercept, and the slope-intercept covariance of the fit.

In cases where the peak is well sampled, we can use the measured peak brightness as a basis for comparison. We consider the peak magnitude to be measured if we have a photometry point within one day of the estimated peak date. Note that, in most cases where we have data, the extrapolated peak is brighter than the observed peak. This is expected since we are fitting a straight line to a concave-down portion of the light curve. Since the maximum difference between estimated and observed peak magnitude is 0.27~mag, we prefer to accept this bias rather than using a more sophisticated method to fit the well-sampled peaks.

The decline slope used for extrapolation, the estimated peak magnitude, and the observed peak magnitude for each supernova are listed in Table~\ref{tab:lcparams}.

\subsubsection{Explosion Epoch and Rise Time}
In order to determine the rise time, we must estimate the epoch of explosion. Where possible, we fit a parabola (flux vs.\ phase) to the premaximum light curve. We use the zero of this fit as our estimate of the explosion time. For objects not observed before maximum light, it is impossible to independently estimate the explosion epoch and the peak date. For those objects, we place an upper limit on the rise time that corresponds to the time between the last nondetection and the first detection. The last nondetections of SN~2005la and ASASSN-15ed are not deep enough to constrain the epoch of explosion, so we do not estimate their rise times. See Appendix~\ref{ap:explosion} for per-object details on this estimate. We then calculate the rise time as the time between our estimated explosion date and our estimated peak date in the rest frame of the supernova. The estimated explosion epoch and implied rise times for each object are listed in Table~\ref{tab:lcparams}.

\subsubsection{Results}\label{sec:photresults}
\begin{figure}
\includegraphics[width=\columnwidth]{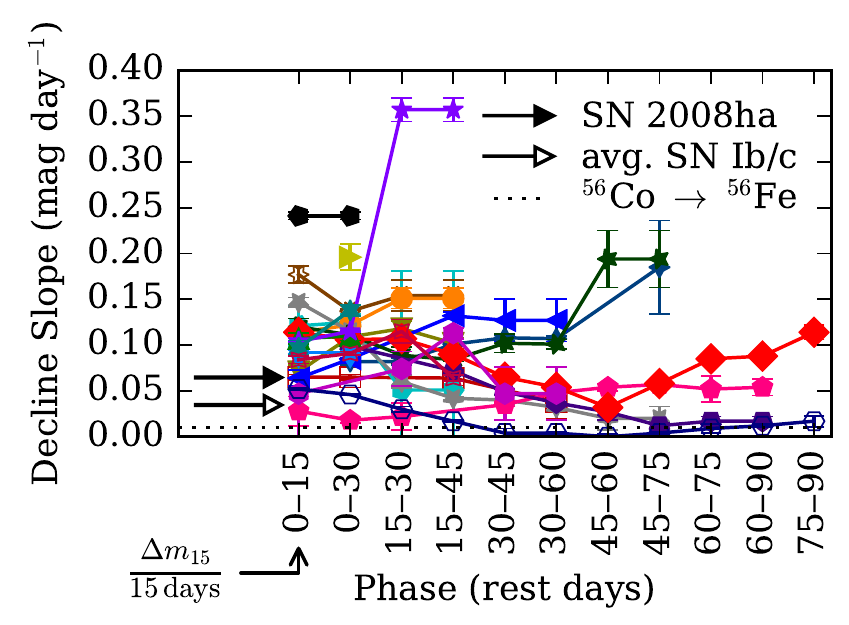}
\caption{Slopes of the post-maximum light curves. Three comparisons are provided: $\Delta m_{15}(R) = 0.97$ for the 02cx-like SN~2008ha \citep[black arrow]{Foley2009}, the weighted average $\Delta m_{15}(r') = 0.52$~mag of the stripped-envelope supernovae in \citet[white arrow]{Bianco2014}, and the $^{56}$Co to $^{56}$Fe decay rate (0.01~mag~day$^{-1}$, dashed line). Note the clustering around 0.10~mag~day$^{-1}$ during the first month after peak, faster than all the comparisons. This may suggest that SNe~Ibn are powered by a short period of circumstellar interaction and/or that their $^{56}$Ni production and explosion energy are smaller than those of other hydrogen-poor supernovae \citep[but see Section~\ref{sec:photresults}]{Moriya2016}.}
\label{fig:declineslope}
\end{figure}

The most striking result of the photometric analysis is the relative homogeneity in the shapes of Type~Ibn light curves. Whereas Type~IIn light curves can decline in a few weeks or last for many months \citep[see, e.g.,][their Figures~17 and 18]{Kiewe2012}, almost all Type~Ibn light curves decline at rates of 0.05--0.15~mag~day$^{-1}$ during the first month after maximum light (see Figure~\ref{fig:declineslope}). These fast decline rates rule out radioactive decay as a significant power source for the late-time light curves of all but the most extreme outlier (OGLE12-006). We are therefore in agreement with the rest of the literature that circumstellar interaction most likely powers SNe~Ibn.

However, since one might expect interaction-powered light curves to be as diverse as the set of possible CSM density profiles that generate them, our data suggest a relatively homogeneous set of progenitor systems and/or explosion parameters. \cite{Moriya2016} explain this homogeneity by proposing that the duration of circumstellar interaction is shorter for SNe~Ibn than for Type~IIn supernovae, despite similar explosion properties and CSM configurations. Our photometry also supports this hypothesis. Our sample of spectra, on the other hand, continue to show evidence of interaction several weeks after maximum light. In particular, spectra of SNe~Ibn show narrow emission lines and are bluer than those of other hydrogen-poor supernovae well over a month after maximum light (see bottom panel of Figure~\ref{fig:comparespec}). Together, the light curves and spectra seem to indicate decreased interaction at later times: strong enough to produce narrow lines and blue spectra but weak enough not to significantly affect the light curves.

\begin{figure}
\includegraphics[width=\columnwidth]{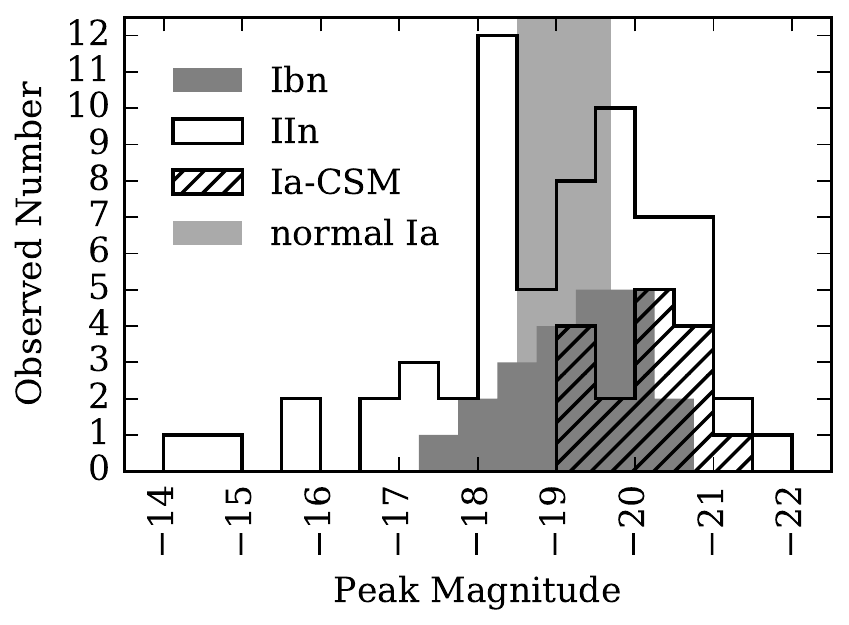}
\caption{Observed distribution of the estimated peak absolute magnitudes ($R$- or $r$-band, except as noted in Table~\ref{tab:lcparams}; see Section~\ref{sec:decline_peakmag}) of SNe~Ibn compared to those of Type~IIn, Type~Ia-CSM, and normal SNe~Ia (all in the $r$-band), all of which are relatively luminous compared to normal CCSNe. Data for Type~IIn and Ia-CSM supernovae are from \cite{Silverman2013}. The Type~Ibn distribution appears narrower than the Type~IIn distribution, but the difference is not statistically significant.}
\label{fig:hist}
\end{figure}

Peak magnitudes of Type~IIn supernovae have a wider observed distribution than those of SNe~Ibn (see Figure~\ref{fig:hist}). However, with only 22 objects, we are unable to rule out the possibility that they arise from the same underlying distribution. We perform the Kolmogorov--Smirnov test, comparing the sample distributions of Type~IIn and Ibn peak magnitudes, which yields a statistically insignificant result: the maximum distance between the cumulative distribution functions of the two samples is $D=0.23$ with $p=0.32$. Furthermore, the faster decline rates of SNe~Ibn might cause an observational bias against discovering fainter objects, making the underlying distributions more similar.

\begin{figure*}
\centering
\includegraphics[width=\textwidth]{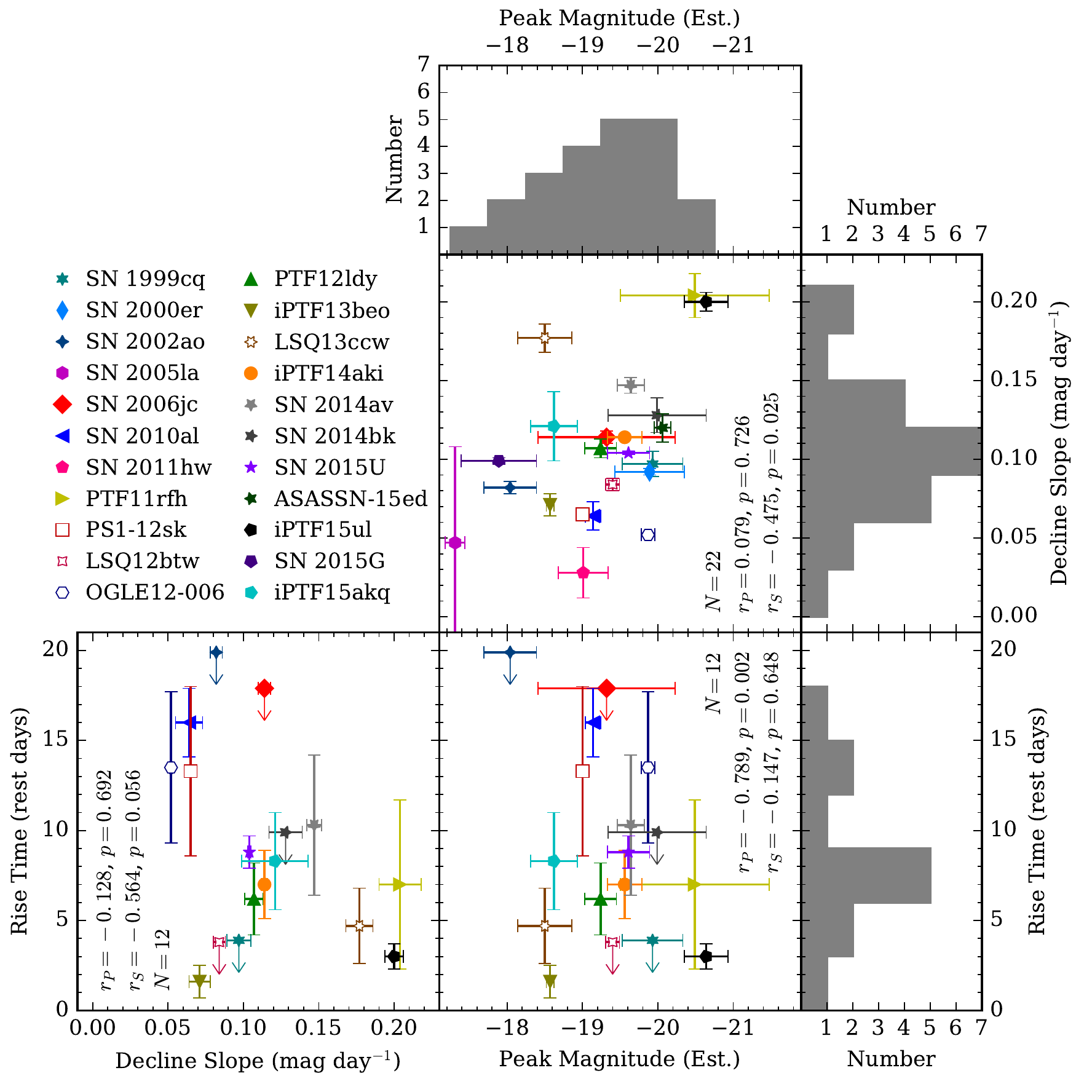}
\caption{Histograms and scatter plots of light curve parameters. The weighted Pearson correlation coefficient ($r_P$), the Spearman rank correlation coefficient ($r_S$), and the associated $p$-values (probability of chance correlation) are given for each pair of parameters. Histograms and statistics exclude points with rise-time upper limits. For consistency, the estimated peak magnitude (see Section~\ref{sec:decline_peakmag}) is plotted for all objects, regardless of whether their peaks were well observed. We do not find any correlations that are statistically significant in both $r_P$ and $r_S$.}
\label{fig:correl}
\end{figure*}

Finally, we search for correlations between the three parameters we estimate: decline slope, peak magnitude, and rise time. For each pair of parameters, we calculate the weighted Pearson correlation coefficient ($r_P$, which down-weights points with large error bars but assumes the variables have Gaussian distributions), the Spearman rank correlation coefficient ($r_S$, which does not assume Gaussian distributions but weights all points equally), and their associated $p$-values (see Figure~\ref{fig:correl}). We do not find any correlations that are statistically significant in both $r_P$ and $r_S$.

In the figures throughout this paper, objects in the emission subclass are assigned cool colors (green/blue/purple) while objects in the P~Cygni subclass are assigned warm colors (pink/red/orange/yellow). Objects not obviously belonging to either class are colored gray and black. Since warm and cool colors appear to be well mixed in each plot, we conclude that light curve properties do not correlate strongly with the two spectral subclasses.

\section{Light Curve Templates}
\begin{turnpage}

\defcitealias{Pastorello8}{P15d}
\defcitealias{Tsvetkov2015}{T15}
\defcitealias{Shivvers2016}{S16}

\setlength{\tabcolsep}{2pt}
\begin{deluxetable*}{llccccccccl}
\tablecaption{Light Curve Parameters}
\tablehead{\colhead{Name} & \colhead{Filter} & \colhead{Non-Det.} & \colhead{Explosion} & \colhead{First Det.} & \colhead{Peak Date} & \colhead{Rise Time} & \colhead{Decline Slope} & \colhead{Peak Mag.} & \colhead{Peak Mag.} & \colhead{Data References} \\
& & \colhead{(MJD)} & \colhead{(MJD)} & \colhead{(MJD)} & \colhead{(MJD)} & \colhead{(days)} & \colhead{(mag day$^{-1}$)} & \colhead{(Est.)} & \colhead{(Obs.)} &}
\startdata
SN 1999cq & R & 51344.4\phn & \nodata & 51348.4\phn & $51348 \pm 3$\hf  & $< 3.9$ & $0.097 \pm 0.008$ & $-19.93 \pm 0.40$\hf & $-19.73 \pm 0.10$\hf & \cite{Matheson2000}     \\
SN 2000er & R & 51824.96 & \nodata & 51873.18   & $51869 \pm 5$\hf  & $< 46.6$ & $0.092 \pm 0.002$ & $-19.89 \pm 0.46$\hf & \nodata & \cite{Pastorello1}  \\
SN 2002ao & R & 52278.9\phn & \nodata & 52312.24   & $52283 \pm 4$\hf & $< 19.9$ & $0.082 \pm 0.004$ & $-18.04 \pm 0.35$\hf & \nodata & \cite{Foley2007}, \cite{Pastorello1}  \\
SN 2005la & R & 53694.48 & \nodata  & 53695.49   & $53694.8 \pm 1.5$ & \tablenotemark{a} & $0.047 \pm 0.061$ & $-17.31 \pm 0.13$\hf & $-17.28 \pm 0.08$\hf & \cite{Pastorello2}  \\
SN 2006jc & R & 53999.80 & \nodata & 54017.75   & $54008 \pm 8$\hf & $< 17.9$ & $0.114 \pm 0.004$ & $-19.32 \pm 0.91$\hf & \nodata & \cite{Foley2007}, \cite{Pastorello2007,Pastorello1} \\
SN 2010al & R & 55267.71 & $55267.5 \pm 1.5$\hf       & 55268.03   & $55283.8 \pm 1.1$ & $16.0 \pm 1.9$ & $0.064 \pm 0.009$ & $-19.14 \pm 0.10$\hf & $-18.87 \pm 0.09$\hf & \cite{Pastorello4}  \\
SN 2011hw & R & 55542.\phn\phn & \nodata & 55883.72   & $55874 \pm 10$\hf       & $< 334$ & $0.028 \pm 0.016$ & $-19.01 \pm 0.33$\hf & \nodata & \cite{Smith2012}, \cite{Pastorello4} \\
PTF11rfh  & R, r & 55902.11 & $55902.6 \pm 0.5$\hf                     & 55903.10   & $55911 \pm 5$\hf & $7.0 \pm 4.7$ & $0.204 \pm 0.014$ & $-20.49 \pm 0.99$\hf & \nodata & this work       \\
PS1-12sk  & z\tablenotemark{b} & 55985.3\phn & $55992.0 \pm 5.0$\hf           & 55997.3\phn & $56006.1 \pm 0.3$  & $13.3 \pm 4.7$ & $0.065 \pm 0.003$ & $-19.00 \pm 0.03$\hf & $-19.05 \pm 0.02$\hf & \cite{Sanders2013}      \\
LSQ12btw  & g & 56009.12 & $56009.5 \pm 1.0$\hf & 56013.11   & $56013.1 \pm 1.0$ & $< 3.8$ & $0.084 \pm 0.004$ & $-19.40 \pm 0.09$\hf & $-19.44 \pm 0.04$\hf & \cite{Pastorello6}  \\
OGLE12-006 & I & 56199.26 & $56203.3 \pm 4.0$\hf       & 56207.34   & $56217.6 \pm 1.8$ & $13.5 \pm 4.2$ & $0.052 \pm 0.001$ & $-19.87 \pm 0.09$\hf & $-19.82 \pm 0.02$\hf & \cite{Pastorello5}  \\
PTF12ldy  & R\tablenotemark{c} & 56235.14 & $56236.1 \pm 1.0$\hf          & 56237.12   & $56243 \pm 2$\hf        & $6.2 \pm 2.0$ & $0.107 \pm 0.006$ & $-19.24 \pm 0.21$\hf & $-19.20 \pm 0.02$\hf & this work       \\
iPTF13beo & R & 56428.85 & $56431.2 \pm 0.1$        & 56431.35   & $56433.0 \pm 1.0$ & $1.6 \pm 0.9$ & $0.071 \pm 0.007$ & \tablenotemark{d} & $-18.57 \pm 0.05$\hf & \cite{Gorbikov2014}     \\
LSQ13ccw  & g & 56533.10 & $56534.0 \pm 1.0$\hf       & 56535.02   & $56539 \pm 2$\hf       & $4.7 \pm 2.1$ & $0.177 \pm 0.009$ & $-18.50 \pm 0.36$\hf & $-18.46 \pm 0.06$\hf & \cite{Pastorello6}  \\
iPTF14aki     & R & 56756.45 & $56756.5 \pm 0.3$\hf       & 56757.42   & $56764 \pm 2$\hf       & $7.0 \pm 1.9$ & $0.114 \pm 0.002$ & $-19.56 \pm 0.23$\hf & $-19.30 \pm 0.03$\hf & this work \\
SN 2014av     & R & 56753.81 & $56760.0 \pm 3.8$\hf & 56763.84 & $56770.6 \pm 1.2$ & $10.3 \pm 3.9$ & $0.147 \pm 0.005$ & $-19.64 \pm 0.18$\hf & $-19.67 \pm 0.10$\hf & \cite{Pastorello9} \\
SN 2014bk     & R\tablenotemark{e} & 56802.50 & \nodata & 56805.51 & $56808 \pm 5$\hf & $< 9.9$ & $0.128 \pm 0.011$ & $-19.99 \pm 0.65$\hf & \nodata & \cite{Pastorello9} \\
SN 2015U     & r & 57034.99 & $57062.6 \pm 0.4$\hf & 57062.97 & $57071.5 \pm 0.8$\hf & $8.8 \pm 0.9$ & $0.104 \pm 0.001$ & $-19.61 \pm 0.28$\tablenotemark{f} & $-19.41 \pm 0.27$\tablenotemark{f} & \citetalias{Tsvetkov2015}, \citetalias{Pastorello8}, \citetalias{Shivvers2016}, this work\tablenotemark{g} \\
ASASSN-15ed     & r\tablenotemark{e} & 57078.57 & \nodata & 57082.59 & $57086.9 \pm 0.6$ & \tablenotemark{a} & $0.120 \pm 0.009$ & $-20.06 \pm 0.11$\hf & $-19.90 \pm 0.25$\hf & \cite{Pastorello7} \\
iPTF15ul & R, r\tablenotemark{h} & 57090.38 & $57090.7 \pm 0.7$\hf & 57091.38 & $57093.9 \pm 0.5$ & $3.0 \pm 0.7$ & $0.200 \pm 0.006$ & $-20.64 \pm 0.29$\tablenotemark{f} & $-20.43 \pm 0.26$\tablenotemark{f} & this work \\
SN 2015G & r & 57034.45 & \nodata & 57104.79 & $57100. \pm 5$\hf & $< 70$ & $0.099 \pm 0.002$ & $-17.89 \pm 0.50$\hf & \nodata & this work \\
iPTF15akq & R, r & 57125.26 & $57124.8 \pm 1.7$\hf & 57126.50 & $57134.0 \pm 2.5$ & $8.3 \pm 2.7$ & $0.121 \pm 0.022$ & $-18.62 \pm 0.31$\hf & \nodata & this work
\enddata
\tablecomments{
\tablenotetext{a}{The nondetection is not deep enough to constrain the explosion epoch.}
\tablenotetext{b}{Last nondetection in $y$-band.}
\tablenotetext{c}{Some $g$-band points converted to $R$-band using $g-R = -0.16$, the average color of iPTF14aki in the first week after peak.}
\tablenotetext{d}{The decline from the first peak was not observed, so no peak magnitude can be generated using our method. The observed magnitude of the first peak is used in the remaining analysis.}
\tablenotetext{e}{Last nondetection and first detection in $V$-band.}
\tablenotetext{f}{Error on peak absolute magnitude includes error from extinction estimate.}
\tablenotetext{g}{Abbreviations: \citetalias{Tsvetkov2015} = \cite{Tsvetkov2015}; \citetalias{Pastorello8} = \cite{Pastorello8}; \citetalias{Shivvers2016} = \cite{Shivvers2016}.}
\tablenotetext{h}{Last nondetection and first detection in $g$-band.}
\label{tab:lcparams}}
\end{deluxetable*}

\end{turnpage}

\begin{turnpage}

\setlength{\tabcolsep}{3pt}
\begin{deluxetable*}{lccccccccccc}
\tablecaption{Decline Slopes (mag day$^{-1}$)}
\tablehead{\colhead{Name} & \colhead{$0-15$ days} & \colhead{$0-30$ days} & \colhead{$15-30$ days} & \colhead{$15-45$ days} & \colhead{$30-45$ days} & \colhead{$30-60$ days} & \colhead{$45-60$ days} & \colhead{$45-75$ days} & \colhead{$60-75$ days} & \colhead{$60-90$ days} & \colhead{$75-90$ days}}
\startdata
SN 1999cq & $0.097 \pm 0.008$ & $0.138 \pm 0.006$ & \nodata & \nodata & \nodata & \nodata & \nodata & \nodata & \nodata & \nodata & \nodata \\
SN 2000er & $0.092 \pm 0.002$ & $0.092 \pm 0.002$ & \nodata & \nodata & \nodata & \nodata & \nodata & \nodata & \nodata & \nodata & \nodata \\
SN 2002ao & \nodata & $0.082 \pm 0.004$ & $0.082 \pm 0.004$ & $0.101 \pm 0.002$ & $0.108 \pm 0.003$ & $0.107 \pm 0.003$ & \nodata & $0.185 \pm 0.051$ & \nodata & \nodata & \nodata \\
SN 2005la\tablenotemark{a} & $0.047 \pm 0.061$ & \nodata & $0.074 \pm 0.019$ & $0.113 \pm 0.006$ & $0.047 \pm 0.029$ & $0.047 \pm 0.029$ & \nodata & \nodata & \nodata & \nodata & \nodata \\
SN 2006jc & $0.114 \pm 0.004$ & $0.106 \pm 0.000$ & $0.107 \pm 0.001$ & $0.090 \pm 0.000$ & $0.065 \pm 0.001$ & $0.054 \pm 0.000$ & $0.032 \pm 0.002$ & $0.058 \pm 0.001$ & $0.085 \pm 0.002$ & $0.088 \pm 0.002$ & $0.114 \pm 0.008$ \\
SN 2010al & $0.064 \pm 0.009$ & $0.085 \pm 0.005$ & \nodata & $0.132 \pm 0.004$ & $0.127 \pm 0.023$ & $0.127 \pm 0.023$ & \nodata & \nodata & \nodata & \nodata & \nodata \\
SN 2011hw & $0.028 \pm 0.016$ & $0.018 \pm 0.003$ & $0.022 \pm 0.015$ & \nodata & $0.035 \pm 0.002$ & $0.048 \pm 0.001$ & $0.054 \pm 0.004$ & $0.057 \pm 0.002$ & $0.052 \pm 0.014$ & $0.054 \pm 0.009$ & \nodata \\
PTF11rfh\tablenotemark{b} & \nodata & $0.196 \pm 0.014$ & \nodata & \nodata & \nodata & \nodata & \nodata & \nodata & \nodata & \nodata & \nodata \\
PS1-12sk\tablenotemark{c} & $0.065 \pm 0.003$ & $0.065 \pm 0.003$ & \nodata & $0.064 \pm 0.008$ & \nodata & $0.038 \pm 0.018$ & \nodata & \nodata & \nodata & \nodata & \nodata \\
LSQ12btw & $0.084 \pm 0.004$ & $0.091 \pm 0.002$ & $0.115 \pm 0.013$ & $0.064 \pm 0.003$ & \nodata & \nodata & \nodata & \nodata & \nodata & \nodata & \nodata \\
OGLE12-006 & $0.052 \pm 0.001$ & $0.046 \pm 0.001$ & $0.030 \pm 0.003$ & $0.017 \pm 0.001$ & $0.004 \pm 0.005$ & $0.004 \pm 0.001$ & $0.000 \pm 0.003$ & $0.004 \pm 0.001$ & $0.009 \pm 0.005$ & $0.012 \pm 0.002$ & $0.017 \pm 0.006$ \\
PTF12ldy & $0.107 \pm 0.006$ & $0.110 \pm 0.005$ & \nodata & \nodata & \nodata & \nodata & \nodata & \nodata & \nodata & \nodata & \nodata \\
iPTF13beo\tablenotemark{d} & $0.071 \pm 0.007$ & $0.109 \pm 0.003$ & $0.118 \pm 0.008$ & $0.102 \pm 0.006$ & \nodata & \nodata & \nodata & \nodata & \nodata & \nodata & \nodata \\
LSQ13ccw & $0.177 \pm 0.009$ & $0.137 \pm 0.003$ & $0.154 \pm 0.017$ & $0.154 \pm 0.017$ & \nodata & \nodata & \nodata & \nodata & \nodata & \nodata & \nodata \\
iPTF14aki & $0.114 \pm 0.002$ & $0.122 \pm 0.001$ & $0.151 \pm 0.011$ & $0.151 \pm 0.011$ & \nodata & \nodata & \nodata & \nodata & \nodata & \nodata & \nodata \\
SN 2014av & $0.147 \pm 0.005$ & $0.115 \pm 0.002$ & $0.060 \pm 0.007$ & $0.042 \pm 0.002$ & $0.040 \pm 0.005$ & $0.032 \pm 0.003$ & $0.020 \pm 0.013$ & $0.020 \pm 0.013$ & \nodata & \nodata & \nodata \\
SN 2014bk & \nodata & $0.128 \pm 0.011$ & \nodata & \nodata & \nodata & \nodata & \nodata & \nodata & \nodata & \nodata & \nodata \\
SN 2015U & $0.104 \pm 0.001$ & $0.115 \pm 0.001$ & $0.357 \pm 0.013$ & $0.357 \pm 0.013$ & \nodata & \nodata & \nodata & \nodata & \nodata & \nodata & \nodata \\
ASASSN-15ed & $0.120 \pm 0.009$ & $0.111 \pm 0.002$ & $0.089 \pm 0.009$ & $0.084 \pm 0.002$ & $0.102 \pm 0.010$ & $0.101 \pm 0.005$ & $0.194 \pm 0.031$ & $0.194 \pm 0.031$ & \nodata & \nodata & \nodata \\
iPTF15ul & $0.241 \pm 0.004$ & $0.241 \pm 0.004$ & \nodata & \nodata & \nodata & \nodata & \nodata & \nodata & \nodata & \nodata & \nodata \\
SN 2015G & \nodata & $0.099 \pm 0.002$ & \nodata & $0.072 \pm 0.001$ & $0.049 \pm 0.003$ & $0.037 \pm 0.001$ & $0.027 \pm 0.009$ & $0.012 \pm 0.003$ & $0.017 \pm 0.005$ & $0.017 \pm 0.005$ & \nodata \\
iPTF15akq & $0.121 \pm 0.022$ & $0.125 \pm 0.009$ & $0.051 \pm 0.130$ & $0.051 \pm 0.130$ & \nodata & \nodata & \nodata & \nodata & \nodata & \nodata & \nodata
\enddata
\tablecomments{
\tablenotetext{a}{To avoid the second rise, the 0--15 day figure only includes the first three detections, and the 0--30 day figure is omitted.}
\tablenotetext{b}{To avoid the second rise, the 15--45 day figure is omitted.}
\tablenotetext{c}{The point $<1$ day before peak is included to improve the fit.}
\tablenotetext{d}{To avoid the second rise, points $<8$ days after peak are excluded.}
\label{tab:declineslope}}
\end{deluxetable*}
\setlength{\tabcolsep}{6pt}

\end{turnpage}

Motivated by the observed homogeneity in Type~Ibn light curves, we set out to define a region in magnitude--phase space that contains $\sim 95\%$ of the photometric points in our sample. We call this the Type~Ibn light curve template. For the purposes of the template, we exclude the extreme outlier light curve of OGLE12-006 and the three double-peaked light curves (iPTF13beo and SNe 2005la and 2011hw), as their light curves may reflect different physical processes that we do not wish to capture.

\begin{figure}
\includegraphics[width=\columnwidth]{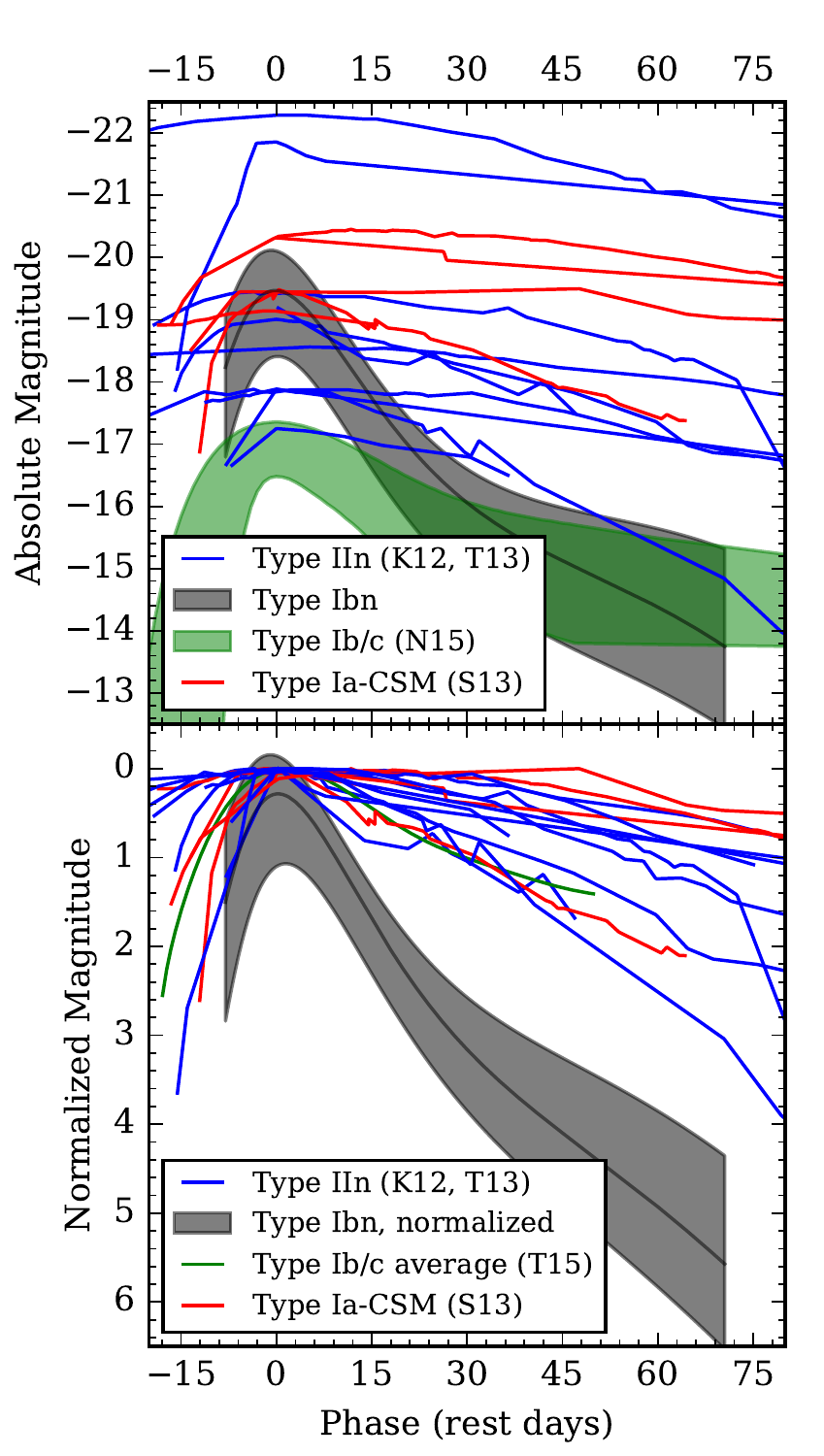}
\caption{Type~Ibn templates compared to analogous Type~Ib/c templates \citep{Nicholl2015,Taddia2015} and samples of Type~IIn \citep[and references therein]{Kiewe2012,Taddia2013} and Type~Ia-CSM \citep[and references therein]{Silverman2013} light curves. The upper panel shows the comparison of absolute magnitudes, and the lower panel illustrates the comparison when all light curves are normalized to peak. SNe~Ibn are much more homogeneous and faster evolving than other interacting supernovae. They are also brighter and faster evolving than non-interacting SNe~Ib/c.}
\label{fig:comparephot}
\end{figure}

First, we use a Gaussian process to fit a smooth curve through the combined light curves of the 18 remaining objects. We perform the fit in log--log space, specifically magnitude vs.\ $\log(\mathrm{phase}+20)$, in order to ensure consistent smoothness between the densely sampled early light curve and the sparsely sampled late light curve. This yields the average Type~Ibn light curve. We then use two more independent Gaussian processes to fit the positive and negative residuals, respectively, from the average light curve. These result in upper and lower $1\sigma$ boundaries for the region. We report the average light curve with $1.96\sigma$ error bars (corresponding to 95\% probability) in Table~\ref{tab:template} and plot the 95\% probability region in the upper panel of Figure~\ref{fig:comparephot}. Although all phases were included in the fit, we only report the template for phases during which more than one supernova was observed. We consider a supernova to be observed during a given phase if its published light curve contains points both before and after that phase. The number of supernovae observed at each phase is reported in the last column of Table~\ref{tab:template}.

Photometric nondetections are completely ignored in the fit, which severely biases the prepeak results. Many objects have extremely short rise times (a few days) that are not contained in the template region, while the few objects with longer rise times pull the fit to brighter magnitudes at early times. When comparing this template to future SNe~Ibn, objects with arbitrarily short rise times should not be considered inconsistent. The purpose of the template is mainly to showcase the fast, homogeneous decline rate and bright peak magnitude.

We also produce a normalized Type~Ibn light curve template by first normalizing all the light curves in our sample to their estimated peak magnitudes and then applying the method described above. Note that although all the normalized light curves pass through 0~mag at 0~days (neglecting the difference between estimated and observed peak magnitudes), the Gaussian process prevents the center and the edges of the template region from passing exactly through this point in order to preserve smoothness. The normalized template is also reported in Table~\ref{tab:template} and appears in the lower panel of Figure~\ref{fig:comparephot}.

\section{Discussion: Implications for the Progenitor}\label{sec:discuss}
Stepping back to Type~Ib supernovae in general, two main progenitor scenarios have been suggested, each with a different way to remove the hydrogen envelope from a massive star. \cite{Gaskell1986}, for example, propose a single massive WR star with a high mass-loss rate to remove the hydrogen envelope before explosion. \cite{Podsiadlowski1992}, on the other hand, propose a close binary system, where binary interaction strips the hydrogen layer from the supernova progenitor. To date, only a single Type~Ib supernova progenitor candidate has been detected: that of iPTF13bvn. In that case, \cite{Cao2013} found that the candidate was consistent with a WR star. However, hydrodynamical modeling by \cite{Fremling2014,Fremling2016} and \cite{Bersten2014} show that the supernova properties are inconsistent with a single massive star and suggest a binary progenitor. A re-analysis of the photometry by \cite{Eldridge2015} also supports the binary hypothesis.

The literature on SNe~Ibn contains many comprehensive discussions of possible progenitor scenarios \citep[see references in Table~\ref{tab:lcparams}, especially][]{Foley2007,Pastorello2007,Pastorello9,Sanders2013}, but they fall roughly along the same lines as for normal SNe~Ib: either binary interaction or stellar winds have stripped the outer layers from a massive star, leaving it surrounded by helium-rich CSM. If the CSM is dense enough, it can interact with supernova ejecta significantly, adding narrow emission lines to the spectra. Although observations of many SNe~Ibn are consistent with single WR progenitors, without a direct progenitor detection, the binary scenario is difficult to rule out. Even the detection of an LBV-like eruption at the location of SN~2006jc two years before explosion could be explained either by an LBV companion or by residual LBV behavior in a recently transitioned WR star \citep{Pastorello2007,Tominaga2008}, although the lack of hydrogen in late-time spectra makes the former scenario less appealing. To complicate things further, \cite{Sanders2013} find that PS1-12sk exploded in a brightest cluster galaxy, and thus may have had a degenerate helium-rich progenitor rather than a massive star. (However, a low-surface-brightness star-forming region might have escaped detection.)

Exploring the dichotomy between spectra that exhibit early P~Cygni lines and those that do not might aid in determining the range of properties of SN~Ibn progenitors. To be clear, we do not suggest that the two subclasses correspond to two entirely different populations of progenitor stars (this makes the late-time similarity hard to explain), but rather that different CSM initial conditions result in at least two varieties of early spectra. We also consider the possibility that \emph{all} SNe~Ibn exhibit helium P~Cygni lines in their spectra immediately after explosion but that our observations thus far have not been sufficiently early to see them for all objects. If this proves to be the case, then the following discussion can be interpreted as explaining why some explosions result in only \emph{short-lived} P~Cygni lines while others show P~Cygni lines for weeks.

One potential explanation for the spectral diversity is optical-depth effects in the CSM. If we attribute the presence of narrow P~Cygni lines to slow-moving, dense circumstellar helium previously ejected by the progenitor, then the optical depth of that material at the time of explosion will determine how it manifests itself in the supernova spectra. If the material is optically thick, it will be backlit by supernova photons. This would result in slow-moving helium features superimposed on a hot thermal continuum. Then as the ejecta sweep up the material, the P~Cygni lines transition into broad supernova features. On the other hand, if the material were optically thin by the time of explosion, broad features would dominate over the CSM emission. Spectra that exhibit very weak helium P~Cygni lines on top of the emission features, such as those of SN~2014av, could represent an intermediate stage, where the circumstellar helium has an optical depth near 1.

\renewcommand{\arraystretch}{2}
\setlength{\tabcolsep}{10pt}
\begin{deluxetable}{cccc}
\tablecaption{Light Curve Templates}
\tablehead{\\[-30pt] \colhead{Phase} & \colhead{Absolute} & \colhead{Normalized} & \colhead{Number of} \\[-7pt] \colhead{(day)} & \colhead{Magnitude} & \colhead{Magnitude} & \colhead{Supernovae}}
\startdata \\[-30pt]
$-8.0$ & $-18.24^{-0.42}_{+0.74}$ & $1.50^{-0.46}_{+0.68}$ & $\phn3$ \\
$-7.5$ & $-18.39^{-0.42}_{+0.71}$ & $1.36^{-0.44}_{+0.66}$ & $\phn3$ \\
$-7.0$ & $-18.53^{-0.41}_{+0.69}$ & $1.22^{-0.42}_{+0.63}$ & $\phn4$ \\
$-6.5$ & $-18.67^{-0.40}_{+0.67}$ & $1.09^{-0.40}_{+0.61}$ & $\phn6$ \\
$-6.0$ & $-18.79^{-0.40}_{+0.65}$ & $0.97^{-0.38}_{+0.59}$ & $\phn7$ \\
$-5.5$ & $-18.91^{-0.39}_{+0.63}$ & $0.85^{-0.36}_{+0.57}$ & $\phn7$ \\
$-5.0$ & $-19.02^{-0.38}_{+0.62}$ & $0.75^{-0.34}_{+0.55}$ & $\phn8$ \\
$-4.5$ & $-19.11^{-0.38}_{+0.61}$ & $0.66^{-0.32}_{+0.53}$ & $\phn8$ \\
$-4.0$ & $-19.19^{-0.37}_{+0.59}$ & $0.58^{-0.31}_{+0.52}$ & $\phn8$ \\
$-3.5$ & $-19.26^{-0.37}_{+0.58}$ & $0.51^{-0.29}_{+0.50}$ & $\phn9$ \\
$-3.0$ & $-19.32^{-0.36}_{+0.57}$ & $0.45^{-0.28}_{+0.49}$ & $\phn9$ \\
$-2.5$ & $-19.37^{-0.35}_{+0.57}$ & $0.40^{-0.27}_{+0.47}$ & $\phn9$ \\
$-2.0$ & $-19.41^{-0.35}_{+0.56}$ & $0.36^{-0.26}_{+0.46}$ & $11$ \\
$-1.5$ & $-19.44^{-0.34}_{+0.55}$ & $0.33^{-0.25}_{+0.45}$ & $11$ \\
$-1.0$ & $-19.46^{-0.34}_{+0.55}$ & $0.31^{-0.24}_{+0.44}$ & $11$ \\
$-0.5$ & $-19.47^{-0.33}_{+0.54}$ & $0.29^{-0.23}_{+0.43}$ & $11$ \\
$\phantom{-}0.0$ & $-19.47^{-0.32}_{+0.54}$ & $0.28^{-0.22}_{+0.42}$ & $11$
\enddata
\tablecomments{The last column lists the number of supernovae included in the fit with observations both before and after the given phase. Results are not reported for phases where only a single supernova has been observed.\\(This table is available in its entirety in machine-readable form.)}
\label{tab:template}
\end{deluxetable}
\renewcommand{\arraystretch}{1}

A second option is that a viewing-angle effect controls whether P~Cygni lines are visible in early Type~Ibn spectra. If the CSM is not spherically symmetric about the progenitor---imagine, for example, a torus of circumstellar helium---then P~Cygni lines may only be visible if the torus is viewed edge-on, whereas emission lines are produced when the torus is face-on. For this scenario to be successful, it must be able to account for the relative frequency of P~Cygni to emission events. This ratio is difficult to estimate, since there are only a small number of events in each class and the classification is not straightforward without early spectra. Nonetheless, with roughly equal numbers in each class (see Table~\ref{tab:disc}), the proposed torus would seem to occupy a large portion of the progenitor star's solid angle.

The uniformity and rapid declines of Type~Ibn light curves could indicate they are powered by interaction with a spatially confined shell of circumstellar helium surrounded by a less dense region of CSM. If the shell is thin enough, strong interaction between the ejecta and the CSM would last only a short time, rather than continuing to power the light curve for many months, as in SNe~IIn. \cite{Quataert2016} suggest that super-Eddington winds tend to produce a CSM shell when driven by energy deposited near the stellar surface, either in the course of binary interaction or by unstable fusion or wave heating in a single star. The wind speed increases as the energy deposition radius decreases, so as the star loses material and shrinks, the ejected mass begins to pile up in a shell around the star. In fact, SN~2014C, which evolved from a hydrogen-poor SN~Ib into a hydrogen-interacting SN~IIn over the course of several months, could represent an extreme case where a massive star ejects its entire hydrogen envelope into a dense shell decades to centuries before explosion. Ejecta from SN~2014C then had to travel through a low-density bubble (during which time it was hydrogen-poor) before reaching and interacting with the circumstellar hydrogen \citep{Milisavljevic2015,Margutti2016}. WR stars and massive binaries are promising candidates for this type of behavior.

Unfortunately, none of these proposals can definitively differentiate between a binary progenitor and a single WR star. For nearby objects, X-ray and radio observations can help directly constrain the CSM properties and provide clues about progenitor mass loss \citep[e.g.,][]{Chevalier2006}. For the rest, early (pre-maximum or first-week-post-maximum) spectroscopy and detailed modeling of the interaction between supernova ejecta and various CSM configurations will be essential to resolving the remaining unknowns.

\section{Summary}
We have presented photometry and spectroscopy of six new SNe~Ibn: PTF11rfh, PTF12ldy, iPTF14aki, iPTF15ul, SN~2015G, and iPTF15akq. We find PTF11rfh and iPTF14aki to be nearly identical to the archetype of the Ibn class, SN~2006jc. Helium P~Cygni lines in early spectra of PTF12ldy link it to SNe~2000er and 2010al. Spectra of iPTF15ul are heavily contaminated by host-galaxy light, but nonetheless resemble those of SN~2000er. Late-time optical spectra of SN~2015G also show similarities to spectra of SN~2010al. iPTF15akq exhibits significant hydrogen in its spectra, making it resemble the transitional Type~Ibn/IIn SN 2005la. When added to objects from the literature, these new events result in a sample of 22 SNe~Ibn. We also presented new data on SN~2015U, previously discussed by \cite{Tsvetkov2015}, \cite{Pastorello8}, and \cite{Shivvers2016}, including a near-infrared spectrum quite similar to that of SN~2010al.

We analyzed the full sample of Type~Ibn light curves and spectra in order to determine the properties of this rare class of objects. Unlike the more commonly observed SNe~IIn, whose interaction with hydrogen-rich CSM has been shown to generate a wide variety of light curve shapes, light curves of SNe~Ibn are more homogeneous and faster evolving, with decline rates clustered closely around 0.1~mag~day$^{-1}$ during the first month after maximum light. We also find that, in the first week after maximum light, two types of Type~Ibn spectra exist: those that consist of helium P~Cygni lines superimposed on a blue continuum and those with broader features that resemble spectra of SN~2006jc.

We hypothesize that both the light curve uniformity and the transition in some objects from narrow P~Cygni lines to broader features can be explained by the presence of a thin shell of circumstellar helium around the progenitor star surrounded by a less dense region of CSM. An optically thick shell backlit by the explosion results in narrow P~Cygni lines that eventually transition into broad emission lines as the shell is swept up by the supernova ejecta. Spectra without P~Cygni lines imply that any previously ejected shell be optically thin at the time of explosion. Alternatively, the spectral diversity might be explained by a viewing-angle effect.

Last, we strongly emphasize that more theoretical modeling work is needed in order to make progress on the progenitor question for SNe~Ibn. In addition, early ($\lesssim 1$~week after peak brightness) observations of future SNe~Ibn will be crucial to constraining these models.

\section*{Acknowledgments}
This work is based on observations obtained with the 48~inch Samuel Oschin Telescope and the 60~inch telescope at the Palomar Observatory as part of the intermediate Palomar Transient Factory (iPTF) project, a scientific collaboration among the California Institute of Technology, Los Alamos National Laboratory, the University of Wisconsin--Milwaukee, the Oskar Klein Center, the Weizmann Institute of Science, the TANGO Program of the University System of Taiwan, and the Kavli Institute for the Physics and Mathematics of the Universe; the New Technology Telescope, operated by the European Organisation for Astronomical Research in the Southern Hemisphere, Chile, as part of PESSTO, ESO program 191.D-0935(C); the Las Cumbres Observatory Global Telescope Network; both the Nordic Optical Telescope, operated by the Nordic Optical Telescope Scientific Association, and the Telescopio Nazionale Galileo, operated by the Fundaci\'{o}n Galileo Galilei of the Italian Istituto Nazionale di Astrofisica, at the Observatorio del Roque de los Muchachos, La Palma, Spain, of the Instituto de Astrof\'{i}sica de Canarias; the Lick Observatory owned and operated by the University of California; and the W. M. Keck Observatory, which was made possible by the generous financial support of the W. M. Keck Foundation and is operated as a scientific partnership among the California Institute of Technology, the University of California, and the National Aeronautics and Space Administration (NASA). We thank the staffs at all of these observatories for their assistance with the observations.

We thank Lars Bildsten and Matteo Cantiello for useful discussions, and all those whose observations and data reduction contributed to this work.
This research has made use of the NASA/IPAC Extragalactic Database (NED), which is operated by the Jet Propulsion Laboratory, California Institute of Technology, under contract with NASA.
The authors made extensive use of the Astropy package \defcitealias{Astropy}{Astropy Collaboration 2013}\citepalias{Astropy} for data analysis.
Part of this research was carried out at the Jet Propulsion Laboratory, California Institute of Technology, under a contract with NASA.
LANL participation in iPTF was funded by the US Department of Energy as part of the Laboratory Directed Research and Development program.

G.H., D.A.H., and C.M. are supported by the National Science Foundation (NSF) under Grant No.~1313484.
A.P., S.B., and N.E.R. are partially supported by PRIN-INAF 2014 with the project ``Transient Universe: unveiling new types of stellar explosions with PESSTO.''
J.S., C.F., E.K., and F.T. gratefully acknowledge support from the Knut and Alice Wallenberg Foundation. The Oskar Klein Centre is funded by the Swedish Research Council.
M.F., A.G.-Y., and M.S. acknowledge support from the European Union FP7 programme through ERC grant numbers 320360, 307260, and 615929, respectively. A.G.-Y. is also supported by the Quantum Universe I-Core program by the Israeli Committee for Planning and Budgeting and the ISF; by Minerva and ISF grants; by the Weizmann-UK ``making connections'' program; and by Kimmel and YeS awards.
A.C. acknowledges support from NSF CAREER award \#1455090.
M.M.K. acknowledges support from NSF PIRE program grant 1545949.
The supernova research of A.V.F.'s group at UC Berkeley is supported by the Christopher R. Redlich Fund, the TABASGO Foundation, and NSF grant AST--1211916. KAIT and its ongoing operation were made possible by donations from Sun Microsystems, Inc., the Hewlett-Packard Company, AutoScope Corporation, Lick Observatory, the NSF, the University of California, the Sylvia \& Jim Katzman Foundation, and the TABASGO Foundation. Research at Lick Observatory is partially supported by a generous gift from Google.

\appendix
\section{Data Reduction}
\subsection{Photometry}\label{sec:phot_reduce}
P48 images were first preprocessed by the Infrared Processing and Analysis Center \citep[IPAC;][]{IPAC}. P60 images were preprocessed using the PyRAF-based\footnote{\url{http://www.stsci.edu/institute/software_hardware/pyraf}} pipeline of \cite{P60}. Image subtraction was then performed using the pipeline of \cite{SullivanPipeline} with pre-explosion images from P48 and images from after the supernova had faded from P60. Finally, instrumental magnitudes obtained through point-spread function (PSF) fitting were calibrated to observations of the same field by the Sloan Digital Sky Survey (SDSS) Data Release 10 \citep{SDSS-DR10}.

KAIT images were reduced using the pipeline of \cite{Ganeshalingam2010}. PSF fitting was then performed using DAOPHOT \citep{DAOPHOT} from the IDL Astronomy User's Library.\footnote{\url{http://idlastro.gsfc.nasa.gov}} The instrumental magnitudes were calibrated to several nearby stars from SDSS, which were transformed into the \cite{Landolt1983} system using the empirical prescription measured by Robert Lupton.\footnote{\url{http://www.sdss.org/dr7/algorithms/sdssUBVRITransform.html\#Lupton2005}}

LCOGT images were preprocessed using the Observatory Reduction and Acquisition Control Data Reduction pipeline \citep[ORAC-DR;][]{ORAC-DR}. Photometry was then extracted using the PyRAF-based {\tt lcogtsnpipe} pipeline \citep[see Appendix~B of ][]{Valenti2016} to perform PSF fitting and calibration to the AAVSO Photometric All-Sky Survey \citep[APASS;][]{APASS} for $BV$ and SDSS Data Release 8 \citep{SDSS-DR8} for $gri$. For iPTF15ul, which lies very near the center of its host galaxy, SDSS images of the same field were subtracted using the algorithm of \cite{Alard2000}, implemented in HOTPANTS\footnote{\url{http://www.astro.washington.edu/users/becker/v2.0/hotpants.html}}, before extracting photometry.

Data from NTT and NOT were reduced using the QUBA photometry pipeline \citep{Valenti2011}.

CSS data were downloaded from the survey's website\footnote{\url{http://crts.caltech.edu}} as calibrated magnitudes. In order to make these magnitudes consistent with the rest of the $V$-band light curve, we converted them to fluxes, determined the average flux at the supernova location in the observing season prior to the explosion, subtracted the average flux from the three detection epochs, and converted back to magnitudes.

Aperture photometry on {\it Swift} images was measured using the method presented by \cite{Brown2009} but with updated zero points from \cite{Breeveld2010}. For iPTF15ul, host-galaxy flux was also measured with the same aperture after the supernova had faded and subtracted from the supernova photometry. Note that {\it Swift} also observed iPTF15akq on three epochs after peak, which resulted in some marginal detections. However, given that the supernova was measured to have $g \gtrsim 21$~mag in P60 subtracted photometry during that period, we take those detections to be of host galaxy only and do not consider them further.

\subsection{Spectroscopy}\label{sec:spec_reduce}
Pre-2013 LRIS and DEIMOS spectra were reduced using standard IRAF\footnote{The Image Reduction and Analysis Facility (IRAF) is distributed by the National Optical Astronomy Observatories, which are operated by the Association of Universities for Research in Astronomy, Inc., under cooperative agreement with the National Science Foundation. \url{http://iraf.noao.edu}} routines \citep[see, e.g.,][]{Cenko2008}. Sky background emission was subtracted using the algorithm described by \cite{Kelson2003}. We removed atmospheric absorption features using the continuum from spectrophotometric standard stars, from which we also derived a sensitivity function for flux calibration.

Post-2013 LRIS spectra were reduced using LPipe,\footnote{\url{http://www.astro.caltech.edu/~dperley/programs/lpipe.html}} a fully automated, end-to-end, high-level IDL pipeline for processing single-object long-slit and imaging data from LRIS.

FLOYDS spectra were reduced using the PyRAF-based {\tt floydsspec} pipeline.\footnote{\url{https://www.authorea.com/users/598/articles/6566}}

Spectra from EFOSC2 were reduced using the PESSTO pipeline. Previous reductions were released via WISeREP as part of the second PESSTO Spectroscopic Survey Data Release \citep[SSDR2;][]{PESSTO}.

The SpeX spectrum was reduced and calibrated using the publicly available Spextool software \citep{spextool}, and corrections for telluric absorption were performed using the IDL tool {\tt xtellcor} developed by \cite{xtellcorr}.

The first DOLORES spectrum was reduced using the QUBA spectroscopy pipeline \citep{Valenti2011}.

All other spectra were reduced using standard IRAF and IDL routines.

\section{Light Curve Fitting}
\subsection{Estimating the Time of Maximum Light}\label{ap:peaks}
Peak dates for SNe~1999cq, 2000er, 2002ao, and 2006jc are given by \cite{Pastorello1} in their Table~1. However, in all cases, their error bars overlap nondetections or nonmaximal points in the light curve. We therefore adopt their estimates with smaller error bars.\footnote{Where the uncertainties are several days, we ignore the distinction between JD--2400000 and MJD.} The peak date for SN~2005la, given by \cite{Pastorello2}, overlaps a nondetection that is deeper than the brightest detection. We shift their peak estimate by $< 1$~day. We note that, out of necessity, their estimates (and some of ours as well) were based partly on comparing the spectral evolution of these objects to SN~2006jc, even though (1) the peak date of SN~2006jc is not well known and (2) there is not enough evidence to support the assumption that all objects of this class have similar spectral series.

The peaks of SN~2010al, OGLE12-006, SN~2014av, SN~2015U, and ASASSN-15ed are well constrained by observations, so we adopt the peaks and uncertainties given in their respective references.

Of the current sample, SN~2011hw has the weakest constraints on the peak, since its position was not observed for 11 months prior to discovery. \cite{Pastorello4} estimate the explosion epoch for SN~2011hw, again using spectral comparisons, to be $\mathrm{MJD} = 55870 \pm 10$, whereas the first detection by \cite{CBET2011hw} occurred on $\mathrm{MJD} = 55883.72$. As an extremely conservative estimate for the date of the (unobserved) first peak, we use $\mathrm{MJD} = 55874 \pm 10$.

PTF11rfh was not observed for a period of 13 days during which its peak likely occurred. Similarly, PTF12ldy was only observed very sparsely around peak and in a different filter than was used to observe its rise and decline. Since the shapes of their light curves are consistent with that of the more densely sampled iPTF14aki, we derive a peak date by matching the absolute magnitudes of these three objects at similar phases. We note that, by adopting these peak dates, the spectra of these three objects are also consistent at similar phases.

The light curve of PS1-12sk was well sampled during the rising phase, but not immediately after peak. Nevertheless, \cite{Sanders2013} fit their $z$-band light curve with a fifth-order polynomial to obtain a peak date of $\mathrm{MJD} = 56006.1 \pm 0.3$. We adopt that estimate here.

LSQ12btw was not detected on the rise, but \cite{Pastorello6} give an upper limit $\sim 4$ mag below and $\sim 4$ days before the first detection. Since no other objects in this sample rise faster than $\sim 1$~mag~day$^{-1}$ on average, we take the peak to be roughly coincident with the first detection.

LSQ13ccw has very tight constraints on the explosion date, but weaker constraints on the peak \citep{Pastorello6}. However, since various low-order polynomial fits to the light curve are consistent with the brightest detection being near the peak, we adopt that assumption here. Note that both these peaks were estimated using the $g$-band photometry presented by \cite{Pastorello6}, which is transformed from LSQ wide-band images.

iPTF13beo was observed every night during the rise to the first peak, ending with two nights of approximately equal brightness. There was then a gap in observing during the fall from the first peak \citep{Gorbikov2014}. We take the peak date to be between the two points at peak with error bars including both points.

iPTF14aki was observed twice during its rise to peak and then every day or two during its initial decline. Again, since various polynomial fits to the light curve are consistent with the brightest detection being the peak, we adopt that assumption here.

The first detection (in the $g$-band) of SN~2014bk by the Kiso Supernova Survey (KISS) was announced in CBET~3894 \citep{CBET2014bk}. Starting 7.5~days later, \cite{Pastorello9} report four additional epochs of photometry in $UBVR$. Based on the proximity of the KISS nondetection, the peak likely occurred during those 7.5 days, but a post-peak discovery cannot be ruled out. We therefore include the entire period between nondetection and second detection in the peak date error range.

iPTF15ul was well observed around maximum light in the $g$-band by iPTF. Two observations of approximately equal brightness were obtained one day apart, so we estimate the peak to have occurred half way between them, with error bars including both observations.

SN~2015G was already declining at discovery. Our peak estimate of five days before discovery results in a light curve that generally matches the others in the sample and spectra that are consistent with those of SN~2010al at similar phase. We extend the error bars such that they include the first observation.

iPTF15akq was observed on the rise and decline, but not for a period of five days around peak. We take the peak date to be in the middle of those five days and include both adjacent observations in our error range.

\subsection{Estimating the Explosion Epoch}\label{ap:explosion}
The explosion epochs for SN~2010al, SN~2011hw, LSQ12btw, OGLE12-006, iPTF13beo, LSQ13ccw, and SN~2014av were estimated in their respective references, and we adopt those estimates here. For LSQ12btw, we consider the magnitude jump from nondetection to detection implausible if the explosion did not occur toward the beginning of the error range, so we decrease the upper error bar to match the lower error bar (one day). However, since SN~2011hw and LSQ12btw were not observed before peak, we do not use these estimates to calculate the rise times.

PTF11rfh only has two pre-maximum points---not enough to which to fit a polynomial---so we can only constrain the explosion epoch to fall between the last nondetection and the first detection. However, because in this case the first detection is not at peak brightness, we do use this estimate to calculate the rise time.

To estimate the explosion epochs of PS1-12sk, SN~2015U, and iPTF15ul, we fit parabolas to the premaximum light curves and find where the fits cross zero. \cite{Sanders2013} alternatively consider the first detection and the last nondetection (in the $y$-band) as the explosion epoch of PS1-12sk, which are 12 days apart. We include most of this range in our error bars. The root of the fit for SN~2015U is only 0.4~days before the first detection, so we adopt this as the uncertainty. Likewise, the root of the fit for iPTF15ul is only 0.7~days before the first detection, so we adopt this as the error.

PTF12ldy was observed three times on the rise: twice in $R$ and once in $g$. We convert the $g$-band point to $R$ using $g-R = -0.16$~mag, which was derived from the light curve of iPTF14aki during the first week after peak. We then fit a parabola to these three points and find that the root is $\mathrm{MJD} = 56236.1 \pm 1.0$, where again the error bar spans the range between the last nondetection and the first detection.

Fitting parabolas to the rising points of iPTF14aki and iPTF15akq give explosion dates that slightly precede their last nondetections. We therefore use those nondetections as estimates of the explosion epochs and adopt errors that are the difference between these and the roots of the fits.

Since we cannot say whether the first detection of SN~2014bk was before or after peak, we do not attempt to estimate the explosion epoch. The upper limit on the rise time is based on the second detection.

\bibliography{library}

\end{document}